\documentclass[12pt]{article}
\usepackage[top=1.0in, bottom=1.0in, left=1.0in, right=1.0in]{geometry}
\usepackage{enumerate,fancyhdr,xcolor}
\usepackage{graphicx}
\usepackage{comment}
\usepackage{hyperref}
\hypersetup{colorlinks=false}
\title{NuLat: \\
A new type of Neutrino Detector for \\
Sterile Neutrino Search at Nuclear Reactors \\and
Nuclear Nonproliferation  Applications}

\begin{document} 
\maketitle
\vskip 0.1in
\begin{center}
C.~Lane$^1$, S.M.~Usman$^2$, J.~Blackmon$^3$, C.~Rasco$^3$, H.P.~Mumm$^4$, D.~Markoff$^5$,  G.R.~Jocher$^6$, R.~Dorrill$^7$,  M.~Duvall$^7$, J.G.~Learned$^7$, V.~Li$^7$, J.~Maricic$^7$, S.~Matsuno$^7$, R.~Milincic$^7$, S.~Negrashov$^7$, M.~Sakai$^7$, M.~Rosen$^7$, G.~Varner$^7$, P.~Huber$^8$, M.L.~Pitt$^8$, S.D.~Rountree$^8$, R.B.~Vogelaar$^8$, T.~Wright$^8$, Z.~Yokley$^8$  \\ 

\vskip 0.07in
\textit{$^1$Drexel University, Department of Physics, 3141 Chestnut St, Philadelphia PA 19104}\\

\vskip 0.07in
\textit{$^2$Johns Hopkins University, Baltimore, MD, 21218}\\

\vskip 0.07in 
\textit{$^3$Louisiana State University, Baton Rouge, LA 70803}\\

\vskip 0.07in
\textit{$^4$National Institute of Standards and Technology, Gaithersburg, MD 20850} \\

\vskip 0.07in
\textit{$^5$North Carolina Central University, 1801 Fayetteville Street, Durham, NC 27707}\\

\vskip 0.07in
\textit{$^6$Ultralytics LLC, 2200 Wilson Blvd, Arlington, VA 22201}\\

\vskip 0.07in
\textit{$^7$University of Hawaii, 2505 Correa Rd, Honolulu, HI, 96822}\\

\vskip 0.1in
\textit{$^8$Center for Neutrino Physics, Virginia Tech, Blacksburg, VA 24061}\\
\end{center}
\setcounter{page}{1}


\begin{abstract}
\textit{We describe a new detector, called NuLat, to study electron anti-neutrinos a few meters from a nuclear reactor, and search for anomalous  neutrino oscillations. Such oscillations could be caused by sterile neutrinos, and might  explain the ``Reactor Antineutrino Anomaly''. NuLat, is made possible by a natural synergy between the miniTimeCube and mini-LENS programs described in this paper. It features a ``Raghavan Optical Lattice" (ROL) consisting of 3375 boron or $^6$Li loaded plastic scintillator cubical cells 6.3\,cm (2.500'') on a side. Cell boundaries have a 0.127\,mm (0.005'') air gap, resulting in total internal reflection guiding most of the light down the 3 cardinal directions. The ROL detector technology for NuLat gives excellent spatial and energy resolution and allows for in-depth event topology studies.  These features allow us to discern inverse beta decay (IBD) signals and the putative oscillation pattern, even in the presence of other backgrounds. We discuss here test venues, efficiency, sensitivity and project status.}
\end{abstract}

\tableofcontents

\section{Introduction to NuLat} 
\hskip 0.3in
Sterile neutrinos (whose existence is hinted by several experiments), if found, would inescapably play a revolutionary role in our understanding of particle physics.  The now established fact of neutrino masses, highlights the incomplete  theoretical understanding of the origin  of neutrino mass.  Most models, of which there are many, invoke a heavy (and sterile at ordinary energies) neutrino state for the missing right-handed neutrinos, which in any case do not participate significantly in weak interactions at energies as yet probed (less than a few hundred\,GeV). Definitive
discovery of a sterile neutrino would point to the scale of physics responsible for neutrino mass, provide
clues to the actual mechanism, potentially open a gateway to the dark sector, and be the first particle found
outside the Standard Model. Sterile neutrinos are well motivated over a wide range of parameter space, but
there has been particular emphasis recently on sterile neutrinos in the 1\,eV mass range due to a variety of
experimental hints~\cite{bib:Stereo}.

This White Paper describes a unique project primarily designed to search for evidence of sterile neutrinos, but which possesses excellent potential for nuclear safeguard and other applications. The detector, called
NuLat (short for Neutrino Lattice), is made possible by a natural synergy between the mini-LENS and
miniTimeCube programs. The goal of the mini-LENS (Low-Energy Solar Neutrino Spectroscopy) program has been to develop the
“Raghavan Optical Lattice” (ROL) technology aimed primarily at low-energy solar-neutrino detection. Excellent spatial and
energy resolution of this technology combined with the custom, fast, digitizing electronics of miniTimeCube will
allow a timely short-baseline reactor experiment sensitive to 1\,eV sterile neutrinos to be performed. ROL technology will enable accurate measurement of the positron's energy in inverse beta decay (IBD) events while cleanly rejecting backgrounds via spatial and temporal coincidence of the IBD neutron tagging. Described below are details of the miniTimeCube and LENS programs and their merger.  

The miniTimeCube (mTC) group has constructed a two-liter neutrino detector to demonstrate the power of using fine pixels and fast timing  to reconstruct detailed event topology in the IBD interaction coming from reactor antineutrinos, even in a scintillation medium (previously believed  to be useful for calorimetric measurement only, due to isotropy of emitted scintillation light). The next step in the TimeCube evolution has been projected to be a scale-up from the two-liter detector to a one-ton detector. With the evolution to a m$^3$ scale and near surface location, (necessitated by reactor locations), it was realized that a dominant problem in scaling up from two liters was going to be background rejection, a  major challenge of any low-energy surface neutrino detector, especially in the close vicinity of a nuclear reactor. 

 After extensive discussions and joint studies between the mTC collaboration and  LENS collaboration, the detector geometry concept of the LENS collaboration has been adopted, as discussed below.  The highly segmented LENS geometry, by which individual cells in a cubical lattice are viewed from six sides, permits three dimensional, digital location of energy deposition within the detector. The unique 3D segmentation of LENS provides superior  background rejection and excellent resolution of the neutrino interaction observed in NuLat simulations.  As such, NuLat is exceptionally well suited to precisely measure reactor antineutrino flux within a few meters from the reactor core, and perform a decisive test of small mixing angle oscillations (at a few percent level) of electron  antineutrinos to a new sterile neutrino species (proposed 4$^{th}$ neutrino generation). Such a breakthrough measurement  has generated a worldwide interest with several experimental pursuits in parallel, to verify or falsify this hypothesis, with high stakes for being first with a clean measurement.

Beyond the search for new physics, NuLat  has potential for multiple applications, ranging from technology demonstration for reactor monitoring purposes, precision (to a few percent) measurement of the antineutrino flux, national security applications, study of an apparent problem with the neutrino spectrum in the range of $5 - 6$\,MeV, and training of young physicists. 

\begin{figure}[!htbp]
\centering
\begin{minipage}{0.45\textwidth}
\includegraphics[scale=0.48]{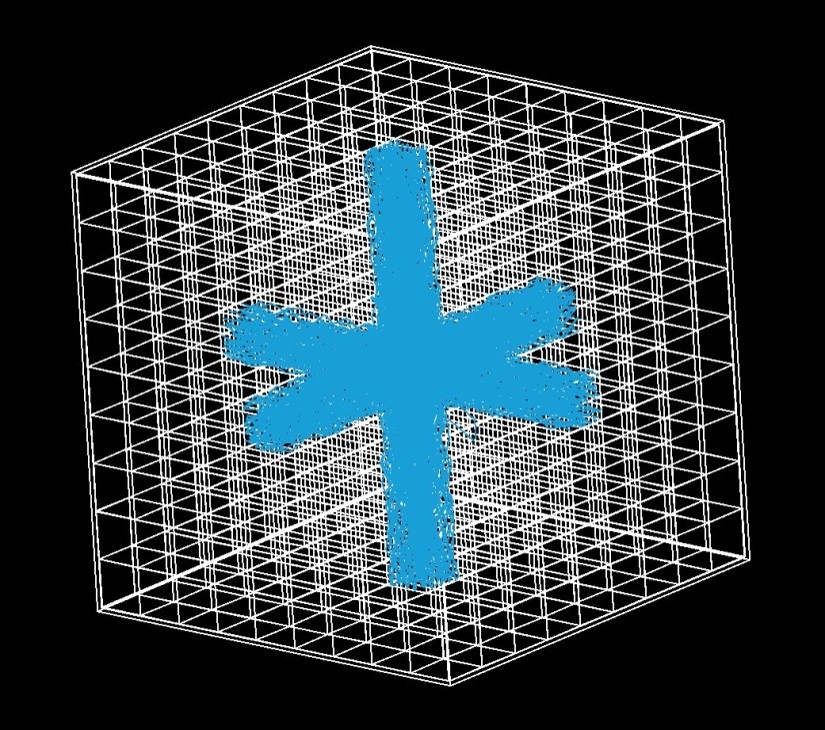}
\caption{Simulation of light channeling in a ROL from an event in the center of the lattice. The blue rays are the tracks of optical photons. Due to the total internal reflection of event light, vast majority of optical photons are channeled along the three principal axes of the detector toward a single photomultiplier tube (PMT) on each of 6 sides of the detector, providing exquisite locating ability. Such efficient channeling leads to well characterized detector response for each cell, allowing for excellent energy reconstruction on a cell by cell basis, and enhanced topology studies for background rejection, including energy deposition sequencing.}
\label{fig:ROL}
\end{minipage}
\hfill
\begin{minipage}{0.5\textwidth}
\includegraphics[scale=0.45]{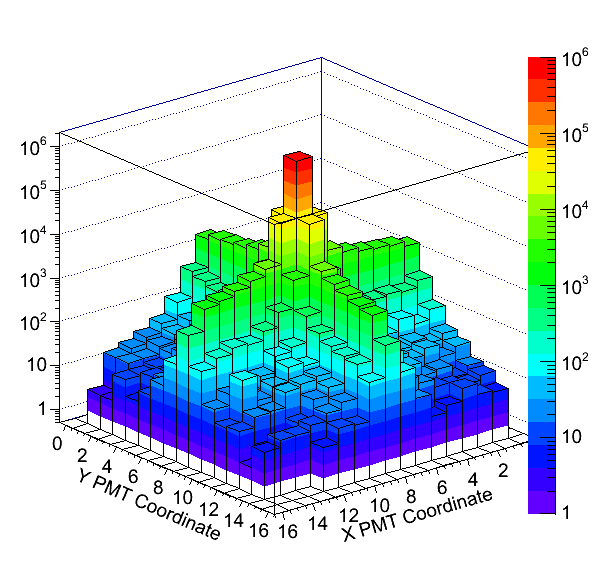} 
		\caption{Log plot of the light output on one face (X-Y) of a mirrored NuLat design due to the deposition of 2\,MeV in the central cell. The amount of light detected in the plane that is not directly facing the cell with the energy deposit is at the level of $<$ 5\%, mostly from dispersion at the opposite mirrored face but a small amount ($<$1\%) from minor in-plane cross talk, and negligible further out channeled light levels $<$5\% of central light. Reducing the enclosing box thickness (6.35\,mm acrylic box (1/4") surrounding the cells) reduces the light dispersion shown here. This pattern is seen in all 3 projections, so that the cube containing the energy deposit is identified uniquely by amplitude alone. Detected light may further be identified by signal timing, permitted location (such as the gammas from positron annihilation must be on average in opposite directions), and sequencing on a per cell basis. This level of spatial and temporal segmentation, in addition to energy resolution, allows for elimination of background.} 
		\label{fig:XYface}
\end{minipage}
\end{figure}

First, background information on sterile neutrinos is reviewed, followed by short baseline reactor physics and the Reactor Antineutrino Anomaly (RAA), reactor flux measurements and predictions, anti-proliferation applications, directional sensitivity, and neutron detection applications. Then, in Section~\ref{sec:NuLatDesign}, general and detailed technical discussion of the new and unique ROL design is presented. The design principles are clearly illustrated in the simulation of light channeling in a ROL detector is shown in figure~\ref{fig:ROL}. A simulated light output on one face of the NuLat detector and distribution of light per cell, due to 10\,MeV deposition in the central cell is shown in figure~\ref{fig:XYface}. Similar output is expected on all six sides of the detector. The central cell along the axis from the  event, has a much larger signal than adjacent cells even on a log scale, contributing to the powerful vertex reconstruction and energy localization, while effectively eliminating background gammas as discussed in detail later in the text. 

\section{Sterile neutrinos -- physics motivation and existing experimental evidence} 
\hskip 0.3in
A sterile neutrino is a neutral lepton with no ordinary interactions (except gravity)  other than those induced by mixing. They are present in most extensions of the Standard Model and in principle can have any mass. For example, sterile neutrinos are a natural ingredient of the most popular and appealing mechanism to generate neutrino masses, the Type I seesaw mechanism ~\cite{bib:Moh80}\cite{bib:Sch80}. They have also been shown to play an important role in leptogenesis ~\cite{bib:Fuk86}\cite{bib:Dav08}, and keV-scale sterile neutrinos could provide a warm dark matter candidate ~\cite{bib:Asa05}. The focus of this review is a relatively light sterile neutrino that mixes significantly with ordinary
neutrinos. There have been a number of recent experimental results that appear anomalous in the context of the standard 3-neutrino framework, but which can be explained by a sterile neutrino with a mass around 1\,eV (summarized in a white paper\cite{bib:Aba12} that provides a comprehensive  summary of this topic).  The mentioned  whitepaper\cite{bib:Aba12}, along with the recent report of the Particle
Physics Project Prioritization Panel (P5)~\cite{bib:Pfi14} recommends  construction of a short baseline neutrino oscillation experiment to clearly and unambiguously
confront the hints of a 1\,eV sterile neutrino. 

Inclusion of a new, sterile neutrino $\nu_s$, requires minimal  3+1 neutrino mixing model  in which the sterile neutrino is acting as a perturbation on the standard three flavor mixing. In this picture, the active neutrino flavors $\nu_e$, $\nu_{\mu}$ and $\nu_{\tau}$ are predominantly composed of three light massive neutrinos $\nu_1$, $\nu_2$ and $\nu_3$ with masses $m_1$, $m_2$ and $m_3$. The sterile neutrino would mostly be composed of a heavy neutrino $\nu_4$ with mass $m_4$, so that $m_4 \gg m_1, m_2, m_3$, implying that $\Delta m^2_{new} = \Delta m^2_{41}$ which is of order 1\,eV$^2$.

In the 3+1 neutrino mixing model, the survival probability in the very short baseline neutrino oscillation experiment is given by:

\begin{equation}
P^{new}_{\nu_{\alpha} \rightarrow \nu_{\alpha}} = 1 - \sin^2 2 \theta_{\alpha, \alpha} \sin^2 \frac{1.27 \Delta m^2_{41}[\mathrm{eV}^2] L[\mathrm{m}]} {E[\mathrm{MeV}]}
\end{equation}
where $\alpha = e, \mu, \tau$.

There are pieces of experimental and observational evidence that either favor and disfavor a $\sim$1\,eV
sterile neutrino that mixes with the three active, light neutrinos. While a briefly review is presented here, a more complete review can be found in ~\cite{bib:Aba12}.

\subsection{LSND and MiniBooNE} 
\hskip 0.3in
The first, and individually still most significant, experimental evidence
in favor of a light sterile neutrino is the result of the LSND experiment at Los Alamos ~\cite{bib:Agu01}, where
electron antineutrinos were observed in a pure muon-antineutrino beam. The most straightforward
interpretation of the LSND result is antineutrino oscillation with mass squared difference $\Delta m^2 \sim 1eV^2$. That
value is incompatible with the mass squared differences implied by solar and atmospheric neutrino
oscillations, so a fourth neutrino is needed to account for this result. Results on the invisible decay width
of the Z boson from the CERN LEP collider imply that a light fourth neutrino would have to be sterile (i.e.
it cannot couple to Z boson) if it exists. Support for the LSND result comes from the MiniBooNE
experiment at Fermilab which has reported a 2.8\,$\sigma$ excess of events in the antineutrino mode ~\cite{bib:Agu13} that
is consistent with neutrino oscillations and the LSND antineutrino appearance signal. MiniBooNE also
observes a 3.4$\,\sigma$ excess of events in the neutrino mode data at low energy ~\cite{bib:Agu13}. On the other hand, the
KARMEN experiment \cite{bib:Arm02} saw no evidence of the disappearance of electron neutrinos, although
probing a similar phase space as LSND. Other constraints disfavoring the sterile neutrino come from the
non-observation of muon-neutrino disappearance by accelerator experiments like CDHSW \cite{bib:Dyd84} and
MINOS \cite{bib:Ada11}. Finally, two more recent accelerator neutrino experiments OPERA \cite{bib:Aga13} and
ICARUS \cite{bib:Seg13} reported negative results in a search for electron neutrinos in a muon-neutrino beam from
CNGS. However, these experiments did not test all regions of the relevant parameter space. 

\subsection{Reactor antineutrino anomaly (RAA)}
\hskip 0.3in
For nearly three decades the expected nuclear reactor antineutrino flux has been calculated using a phenomenological model by conversion of the spectra from the thirty effective beta branches. The method relies on the measurement of the spectrum of fission induced electrons accompanying the antineutrino performed at ILL~\cite{bib:ILL1}~\cite{bib:ILL2}~\cite{bib:ILL3}~\cite{bib:ILL4}.

A new calculation of the reactor antineutrino flux was conducted in 2011~\cite{bib:Mueller} in preparation of the high precision $\theta_{13}$ search in the Double Chooz experiment. The new method relied on the detailed knowledge of the decays of thousands of fission products. Inherent uncertainties in this model came from the various transitions that are not experimentally constrained. 

In addition to reducing the overall uncertainty of the expected antineutrino flux, the new calculation revealed a small overall increase in the expected flux at the level of 3.5\%. When combined with small deficits in the detected antineutrino flux observed in near reactor experiments (between 15\,m and 1.5\,km), the deficit between observed and expected reactor antineutrino rate increases to 5.7\% with the mean value of: 
\begin{equation}
\frac{\phi_{measured}}{\phi_{calculated}} = 0.943 \pm 0.023
\end{equation}

This amounts to a nearly 3\,$\sigma$ effect and the effect was named the Reactor Antineutrino Anomaly (RAA) by G. Mention \textit{et al.}~\cite{bib:Men11}.

If the RAA is due to neutrino mixing, it would require additional neutrino species with $\Delta m^2_{new} \ge 1$ eV$^2$ with the mixing amplitude $\sin^2 (2\theta_{new}) \sim 0.115$. The existence of this new massive sterile neutrino would enable oscillations of electron antineutrinos into sterile neutrinos at distances less than $10 - 15$\,m from the reactor core accompanied by  the flat suppression of the electron antineutrino rate and spectrum observed at distances above 15\,m indicated by the RAA.

The RAA represents an independent hint of the existence of the new massive neutrino species potentially discernable at above 15\,m baselines.

\subsection{SAGE and GALLEX source calibrations} 
\hskip 0.3in
Calibrations with radioactive sources of $^{51}$Cr and
$^{37}$Ar, which both decay via electron capture and emit mono-energetic electron neutrinos, were performed
for the radio-chemical solar neutrino experiments based on gallium (SAGE\cite{bib:Abd99}, and GALLEX\cite{bib:Ans95}). In both cases, a deficit of electron neutrinos of $\sim$25\% was observed \cite{bib:Men11} at
distances of a few meters. This result can be explained by a $\sim$1\,eV sterile neutrino, which would allow
some of the electron neutrinos from the source to “disappear” before they can interact. The effect depends
on nuclear matrix elements, but recent measurements of the relevant Gamow-Teller transitions strengths
\cite{bib:Fre11} seem to support the gallium anomaly.

\subsection{Cosmological evidence}
\hskip 0.3in
Cosmological data, mainly from observations of the cosmic
microwave background and large-scale structure, are sensitive to the possible existence of a light sterile neutrino. Cosmology is sensitive to the effective number of neutrino families N$_{eff}$ primarily because energy density in relativistic particles affects directly the universe's expansion rate during the radiation dominated era. This results in sensitivity to light sterile neutrinos in cosmological observables such as the light elemental abundances from big-bang nucleosynthesis (BBN), the cosmic microwave background (CMB) anisotropies, and the large-scale structure (LSS) distribution. Until very recently, a combination of these observables tended to favor a scenario with a fourth sterile neutrino. However, recent Planck data \cite{bib:Ade13} seems to rule out the possibility of a fully thermalized light neutrino spectrum, finding $N_{eff} =  3.30^{+0.51}_{-0.54}$ at the
95\% confidence level.  WMAP also published their limits based on the 7 years of data~\cite{bib:WMAP7} and results vary depending on data sets included ($CMB+BAO+H_{0}$ or  $CMB+BAO+H_{0}$ + selected galaxy cluster abundances (SPT$_{CL}$)): $N_{eff} = 3.71 \pm 0.35$ and $N_{eff} = 3.29 \pm 0.31$.
These results do not completely rule out the existence of a light sterile neutrino, because there are scenarios where sterile neutrinos would not have thermal energy spectra and number densities \cite{bib:Ful14}.

In summary, there are experimental results that appear anomalous (at the $2 - 3$\,$\sigma$ level) in the context
of the standard three-neutrino framework which can be explained by a sterile neutrino with a mass around
1\,eV. On the other hand, there are a number of results that are in conflict with this interpretation, also at
the $2 - 3$\,$\sigma$ level. It should also be noted that the experimental evidence in favor of a light sterile neutrino is
from effects that are purely in count rates. To clearly resolve this situation, new experiments are needed, particularly ones at short baselines that can observe the energy spectrum distortion implied by a light sterile neutrino in addition to the overall rate deficit. 
The most promising venues are experiments close to nuclear reactor cores, with very strong radioactive sources ($^{144}$Ce -- $^{144}$Pr and $^{51}$Cr) or at accelerators. The prospects of different experiments are discussed in detail in~\cite{bib:Lasserre}.  Experiments very close to nuclear reactors and with intense radioactive sources  are likely to be the first to provide data in favor/against the existence of sterile neutrinos. 

\section{Short baseline reactor neutrino physics} 
\subsection{Overview}
\hskip 0.3in
Nuclear reactors are copious electron antineutrino sources in the range below 10\,MeV. A 1\,GW$_{th}$ nuclear plant produces about $\sim 0.6 \times 10^{20}$ $\bar \nu_e/s$ from the beta decays of the reactor fission products originating mostly from $^{235}$U, $^{238}$U, $^{239}$Pu  and $^{241}$Pu. The knowledge of reactor core composition and burn-up level allows calculation of the neutrino spectrum and rate as a function of time. Since the spectrum and flux of reactor neutrinos are fairly well known, they are well suited for the study of the fundamental neutrino properties and neutrino oscillations in particular. In reactor neutrino experiments, the detected antineutrino rate and spectrum are compared with the expected spectrum and rate based on the reactor burn-up level. Detected rate deficit and spectrum shape distortion indicate disappearance of electron antineutrinos, interpreted as antineutrino oscillations. 

Reactor neutrino experiments with very short baseline ($2 - 18$\,m)  can test the $\sim$1\,eV sterile neutrino hypothesis, as no other neutrino flavor oscillation can take place at such a short baseline.   Several such experiments are coming on line in the current and following years.  The complete list (to the best of our knowledge) is given in the Table~\ref{table:exp_list}.

\begin{table}[!htbp]
\begin{tabular}[c]{|c|c|c|c|c|c|c|c|}
\hline 
Project & Ref. & Dopant & Highly & Moving & P$_{th}$ & L(m) & M$_{target}$ \\
 & & & Segmen. & detector & (MW$_{th})$ & & (tons)\\
\hline
\hline
Nucifer/FRA & \cite{bib:Nucifer} & Gd & & & 70 & 7 & 0.75 \\
\hline
Poseidon/RU & \cite{bib:Poseidon} & Gd & & & 100 & $5 - 15$ & 1.5 \\
\hline
Stereo/FRA & \cite{bib:Stereo} & Gd & & X & 50  & $8.8 - 11.2$ & 1.75 \\
\hline
Neutrino 4/RU & \cite{bib:Neutrino4} & Gd & & X & 100  & $6 - 12$ & 2.2 \\
\hline
Hanaro/KO & \cite{bib:Hanaro}  & Gd, $^6$Li & 2D & X & 30 & 6 & 0.5 \\
\hline
DANSS/RU & \cite{bib:DANSS} & Gd & 2D & X & 3000 & $9.7 - 12.2$ & 0.9 \\
\hline
PROSPECT/USA &  \cite{bib:PROSPECT} & Gd, $^6$Li & 2D & & $20 - 120$ & 4 \& 18 & 1 \& 10 \\
\hline
SoLid/UK & \cite{bib:SOLID}& $^6$Li & 2D & & $45 - 80$ & $6 - 8$ & 3 \\
\hline
NuLat/USA & here & $^{10}$B, $^6$Li & 3D  & X & 1500 & $3 - 8$ & 1.0 \\
\hline
\end{tabular}
\caption{The list of proposed short baseline reactor experiments that emphasizes their main detector characteristics\cite{bib:Lil} and experimental parameters \cite{bib:Lasserre}.}
\label{table:exp_list}
\end{table}
  
All detectors listed in Table~\ref{table:exp_list} detect reactor antineutrinos via the IBD reaction $\bar \nu_e + p^+ \rightarrow e^+ + n$ with an energy threshold of 1.8\,MeV.  IBD has a very prominent detection signature: prompt stopping and annihilation of the positron with an electron in the proton rich target medium, followed by a delayed event in which the thermalized neutron is captured on a nucleus with high neutron-capture cross-section that produces a unique taggable signature. Depending on the nucleus and level of loading in the target, the delayed event takes place on average from a couple of microseconds to a couple hundred microseconds later and has a well defined energy associated with: 1) the deexcitation gammas emitted in the neutron capture on gadolinium or hydrogen, 2) the generation/emission of $^7$Li nucleus, alpha, and gamma from neutron capture on $^{10}$B, or 3) alpha and triton generation from neutron capture on $^6$Li. The prompt-delayed events are correlated both spatially and temporally and can be readily identified.

 A key issue for all of these experiments is the ability to discriminate IBD signals from cosmogenic
and reactor induced backgrounds. While IBD correlated signal possesses a very strong discriminatory
power, high rates of energetic gammas and neutrons can mimic the IBD random coincidence. The
reactor core can generate correlated neutron and high-energy gamma backgrounds as well. 

Very short baseline
experiments  have to operate in whatever space is available near the reactor core to preserve the needed
short baseline, resulting in very small to non-existent overburden protection from any cosmic-ray induced
backgrounds and very limited shielding from the reactor core. The backgrounds must be measured in advance and the detector protected with adequate shielding. As explained previously, the NuLat highly segmented detector design gives
excellent spatial and energy resolution allowing in-depth topology studies to discern IBD events in the face
of these backgrounds. Liquid-free (1\% boron or lithium loaded plastic scintillator) compact design allows it to
be placed uniquely close to extremely compact and powerful naval reactors (Table~\ref{table:exp_list}). 

All of the planned experiments have baselines $\sim1 - 10$\,m to allow for detection of distortions in
the antineutrino energy spectrum expected for a $\sim$1\,eV sterile neutrino. To observe this distortion, compact
reactor cores whose dimensions are $< 1 - 2$\,m (such as a naval reactor) are preferred to large commercial reactors due to smearing of the electron antineutrino source that washes out the oscillation pattern in the data.

\begin{figure}[!htbp]
\centering
\includegraphics[width=6.5in]{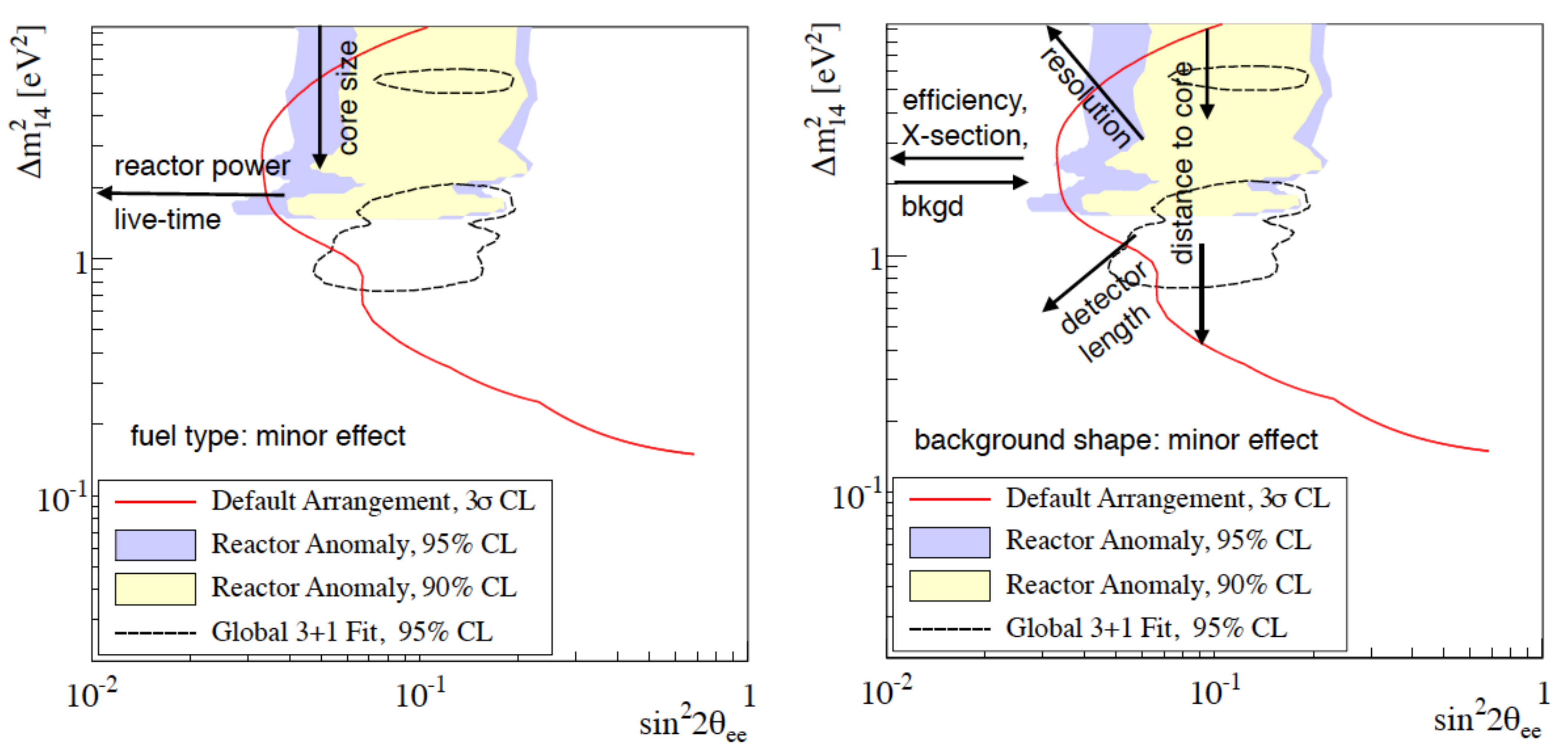}
\caption{Interplay of various detector, source, baseline and statistics effects on the experiment's potential to investigate the sterile neutrino parameter space of interest~\cite{bib:Heeger}. The figure on the left shows how: 1) increase in live-time or reactor power increases sensitivity at smaller mixing angles due to increased statistics of the measurement; 2) an increase in core size decreases sensitivity to larger mixing mass differences, due to smearing of the neutrino baseline associated with a larger core.
The figure on the right illustrates effects associated with the detector and its location with respect to the core as well as available overburden. 1) Increase in detection efficiency and decrease in the cross section uncertainty improves sensitivity to smaller values of mixing angle associated with reduction in the detector associated systematic error
2) Increase in background levels has the opposite effect, decreasing sensitivity to smaller values of mixing angle. This is due to the reduction in the detector’s systematic error. 3) Improvement of energy/vertex resolution increases sensitivity to both higher mixing mass difference and smaller neutrino mixing angle as the detector is more capable to resolve higher frequency oscillations with smaller amplitude, associated with larger squared mass differences and smaller sterile neutrino mixing angles. 4) The distance from the core to the detector shifts the overall sensitivity curve down toward smaller mixing mass differences, with increasing distance, as the oscillations at higher Δm2 are washed out with increase in distance. 5) An increase in detector length improves sensitivity to smaller mixing mass differences and smaller mixing angles.
See~\cite{bib:Heeger} for a comprehensive discussion of how detector parameters affect sensitivity.
}
\label{fig:gen_par_space}
\end{figure}

The impact of short baseline reactor experiments depends on the reactor core’s size and
compactness, background levels, baseline, and detector design optimization. An illustrative plot
demonstrating the interplay of statistics, energy resolutions, and baseline is given in figure~\ref{fig:gen_par_space} \cite{bib:Heeger}.

\subsection{Detailed reactor neutrino flux measurement}
\hskip 0.3in
Precision measurements of reactor neutrino fluxes continue to be of
interest due to a peculiar and unexpected feature observed by the most recent set of reactor neutrino experiments (RENO\cite{bib:RENO}, Double Chooz\cite{bib:DC} and Daya Bay\cite{bib:DB}). All three experiments observed an anomalous,  wide peak at 5\,MeV corresponding to an excess in the number of observed events compared to expectation. The RENO collaboration reported 3.6\,$\sigma$,~\cite{bib:RENO} significance of the excess.  Double Chooz and Daya Bay experiments observed a prominent peak in the same region as well.  The 5\,MeV anomalous peak is not understood as of yet, and it may be due to an unaccounted for part of the reactor flux, but could also be due to a background or new physics. This 5\,MeV peak is indicative of the power of precision neutrino experiment to uncover new phenomena. Resolution of the 5\,MeV anomalous peak requires measurement at a shorter baseline, with a different reactor core composition and with high energy resolution. With its lattice configuration, close to a predominantly $^{235}$U reactor core, NuLat has excellent prospects for understanding this recently revealed puzzle.

\section{Short baseline nonproliferation/national security with NuLat}
\hskip 0.3in
Reactor electron anti-neutrino detection is a promising technique for nuclear \\
non-proliferation control.  An abundant electron anti-neutrino flux is produced by all nuclear reactors and cannot be shielded or falsified. Therefore, monitoring the neutrino flux from a given reactor can verify the operational state of the reactor. For example, anti-neutrino monitoring can reveal if the fuel is changed frequently to enhance Pu production, provide estimate of the power output, and even determine U/Pu fuel mix (with sufficient statistics). 

Monitoring reactors via their neutrino emissions has a long history starting from Reines and Cowan's first detection of neutrinos in the 1950's, followed by a dramatic progress in later decades in Russia.  More recently, the SONGS group deployed a series of detectors at the San Onofre power reactor in Southern California~\cite{bib:LLNL}. Within the last decade, a small worldwide community has emerged, with the common goal of making detectors, which can easily be deployed near (cooperating) reactors, and contribute information to the International Atomic Energy Agency (IAEA) to assist in their treaty obligations to monitor reactors and reactor materials in treaty signatory states around the world. A detector for neutrino monitoring allows taking operations data from the reactor without any interference or risk to the reactor operations. As such, it has a potential to greatly benefit the anti-proliferation goals of the IAEA. Moderate range neutrino monitoring programs have been going ahead in Russia, France, Italy, Germany, England, Brazil, Japan, Korea, and China, as well as the USA. Their progress is reported annually at a series of meetings called AAP (Applied Antineutrino Physics), and is relatively modest.  The detectors report significant backgrounds and require high reactor-off  counts to be subtracted.  Underground detectors work well, as they are shielded from the cosmic rays and the reactor induced backgrounds (neutrons and gammas).  But operating close to the reactor costs doubly. Both reactor and cosmic ray induced backgrounds increase since almost all reactors are at the Earth's surface resulting in a very significant cosmic-ray background.

Several agencies have taken an interest in reactor monitoring in the USA, aside from science funding agencies, for studying neutrino properties.  The Department of Energy NNSA, has funded a series of experiments centered at LLNL.  The NGA, because of interest in mapping world reactors, has funded studies of new types of potential reactor monitoring detectors and the miniTimeCube (mTC) in particular.  The mTC has pioneered the use of very fine pixelization and fast timing to register neutrino interactions (and backgrounds) with unprecedented accuracy (mm scale) and with fast background rejection, on the fly.  The mTC has a fiducial volume of only $\sim$1\,liter (2 liters total), much smaller than other detectors tried or being tested for close in monitoring.  Detector shielding is required at the current  mTC commissioning site close to (2.5\,m) the NIST 20\,MW$_{th}$ to protect against reactor induced background. However, if mTC would be positioned at a large ($3 - 4$\,GW$_{th}$) power reactor, it is designed to operate with \textit{no shielding} while observing 10 neutrino events per day at 20\,m distance. This powerful design feature of mTC is mainly made possible by its outstanding positioning accuracy and powerful background rejection, making the tiny mTC detector a competitive choice for unobtrusive reactor monitoring in cooperative situations.

\subsection{Mobile detector for the monitoring of reactor power and cycles}
\hskip 0.3in
The next step in scaling up the mTC is a meter scale detector.  Such a detector could operate outside the containment facility for a reactor, sitting in a trailer, enabling mobile detector monitoring of nuclear reactors and their power cycles.  Multiplying such a detector by a factor of ten or so, it could still be placed in a shipping container(s) or semi-trailer(s) for mobile operations out to greater distances, at a kilometer scale.  Design of a simple, reliable and affordable meter scale detector, requires R\&D, for which 3D segmentation of NuLat is a promising way forward.  

\subsection{Further Future Long Range Monitoring}
\hskip 0.3in
In 2002, the mTC group began investigating an array of huge neutrino detectors to monitor reactors over large geographical regions since neutrino detection is completely passive and not jam-able.  It is noteworthy that the KamLAND experiment in Japan, at 1000 tons, collected, prior to the 2011 earthquake, neutrinos from reactors all around Japan.  A series of studies were performed on neutrino registration from hundreds to even a thousand kilometers distance and the project was titled NuDAR (\cite{bib:Usman}).  The reason for this is that the well-understood oscillations characteristically distort the neutrino spectrum from the reactor, permitting determination of reactor range from even one detector location. Measurement of neutrino direction or the use of multiple detectors provides additional information to help determine reactor location. Moreover, the characteristic spectral signature of a reactor at a given distance allows one to deconvolve the signals of multiple reactors, or find a small reactor even in the presence of larger reactors in the neighborhood (refer to~\cite{bib:Usman} for further details on this topic).

\section{Neutron detection of Special Nuclear Materials (SNM)}
\hskip 0.3in
A detector such as NuLat has a dual use: in addition to neutrinos, it has  high efficiency for  detecting neutrons due to the neutron capture nuclei loading in the scintillator. This feature allows for the detection of Special Nuclear Materials (SNM) due to their neutron radiation despite rather thick intervening material, such as shipping containers.  Often, two successive scatters in a detector from an energetic neutron can be used to tag the first two scatters of  energetic neutron, consequently extracting the locus of incoming directions. When a neutron in motion is deflected from a straight-line trajectory by scattering from a proton, the angle of the deflection may be calculated given the kinetic energy of the neutron prior to and immediately following the deflection. This simple relationship is expressed by Equation~\ref{eq1}.

\begin{equation}
\sin(\theta/2)  = \sqrt{\frac{E_0 - E_1}{E_0}} = \sqrt{\frac{\Delta E_0}{E_0}}                 
\label{eq1}
\end{equation}

where $\theta$ is the neutron deflection angle, $E_0$ is the original neutron kinetic energy and $E_1$ the neutron kinetic energy after deflection; $\Delta E_0 = E_0 - E_1$  being the difference.  $\Delta E_0$ is directly observable, however $E_0$ is not, and must be inferred via Equation~\ref{eq2}:

\begin{equation}
E_0 = \Delta E_0 + E_1                                                           
\label{eq2}
\end{equation}

where the post-scatter kinetic energy $E_1$ is calculated via the non-relativistic neutron velocity following deflection in Equation~\ref{eq3}:

\begin{equation}
E_1 =\frac{1}{2}m v^2 = \frac{1}{2}\,m \frac{|P_1 - P_2|^2}{(t_2 - t_1)^2}                                            
\label{eq3}
\end{equation}

\begin{figure}[!htbp]
\centering
\includegraphics[width=5.0in]{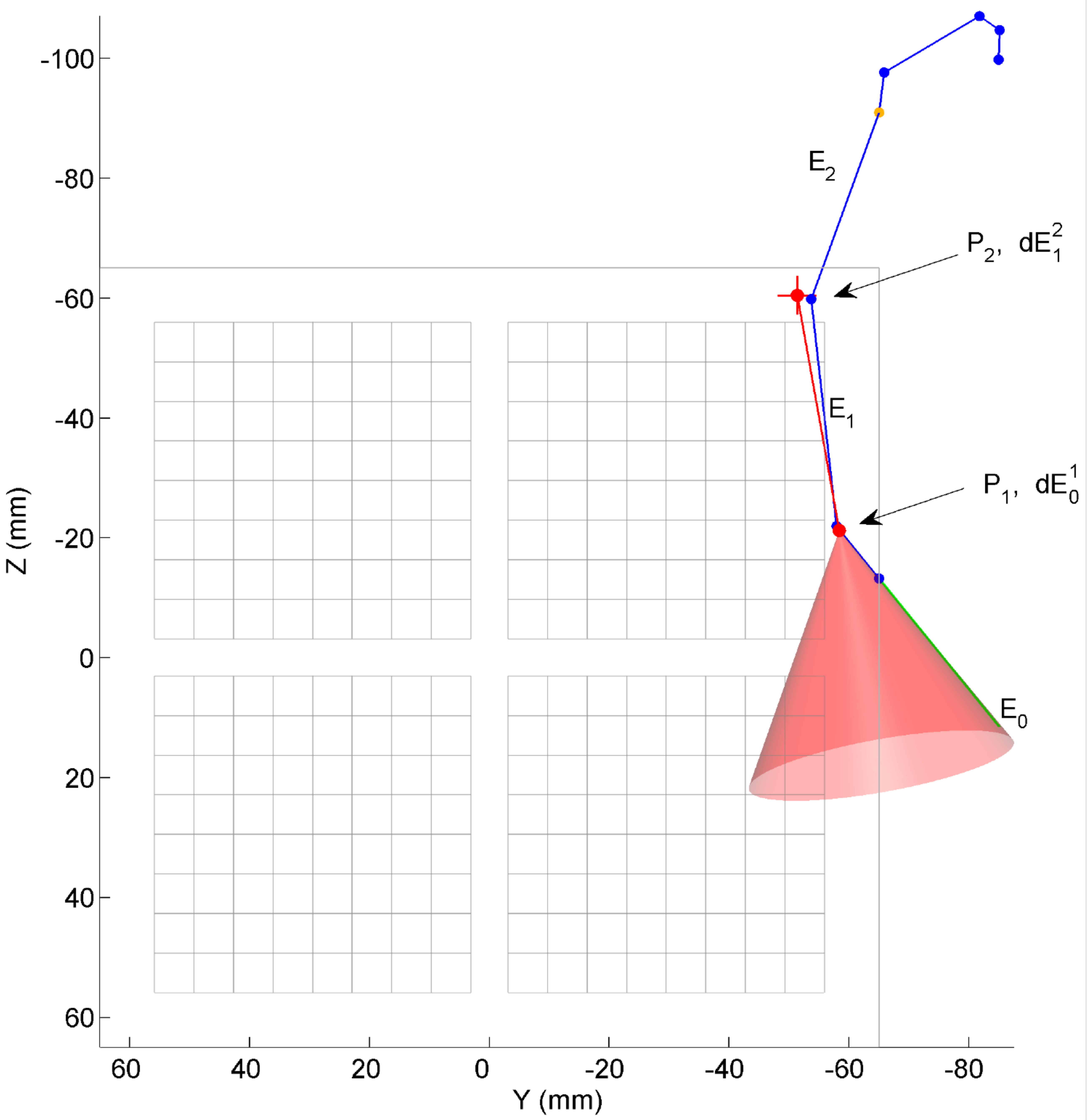}
\caption{Neutron direction estimation diagram. Incoming neutron going upwards is shown in green, while the true travel path through the detector is shown in blue. Estimated travel path is shown in red. Cone about $P_2 - P_1$ vector is  also shown in red.}
\label{fig:nd}
\end{figure}

The times ($t_1$, $t_2$, ...) and positions of neutron scattering events ($P_1$, $P_2$, ...) allow reconstruction of the neutron energy $E_1$, and direction after its first scatter. When combined with the energy deposition $\Delta E_0$ one sets the initial energy $E_0$ and a cone of possible incoming neutron directions.  Figure~\ref{fig:nd} illustrates the described method that leads to determination of the $E_0$ and $\theta$.

\begin{figure}[!htbp]
\centering
\includegraphics[width=5.0in]{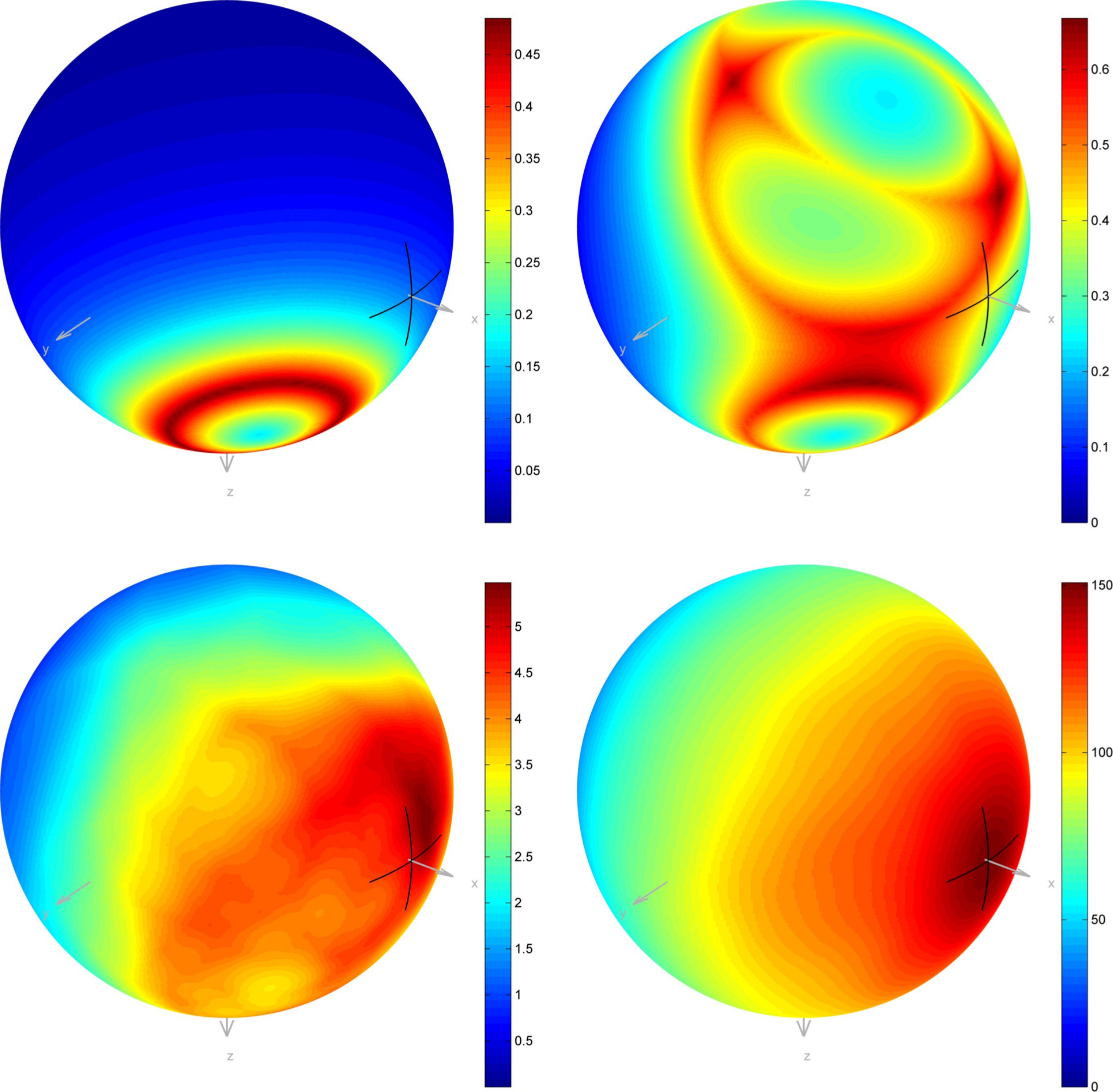}
\caption{Neutron angular probability density functions for 1, 3, 30 and 1000 MC neutrons displayed on unit spheres (going from left to right, in the first and second row.}
\label{fig:sm}
\end{figure}

With this approach, one gets a number of such cones, and makes a ``sky map'' that can be analyzed with Hough Transforms (figure~\ref{fig:sm}) for example. In any event, the surprising fact is that with reasonable resolution one can find sources with useful accuracy, within reasonable times.  As an example, our calculations show that a m$^3$ detector would have good ability to detect a passing truck with a kg of Pu on board.  

Because of homeland security needs, many groups have come up with plans for neutron detectors,  employing various media, densities and geometries.  The detector design described in this white paper has some attractive features due to being solid plastic (no need for $^3$He, not pressurized, nor flammable) and being inherently wide-angle in response and relatively compact.  Future versions will potentially be less expensive as well, particularly in the case of commercial production. In addition, due to the modular nature of the NuLat design the configuration can be optimized based on the deployment of the detector. This white paper focuses on the neutrino detection and background rejection for neutrino detection, though one should keep in mind the potential dual use. The present NuLat represents a new class of such detectors and this development for a short-term physics goal, has long-term importance in other areas.

\section{NuLat Detector Design: The Raghavan Optical Lattice (ROL) }
\hskip 0.3in
It is apparent from these arguments that an above-ground mobile
cubic-meter scale antineutrino detector would be ideal for fundamental physics research (nuclear non-proliferation control and neutron detection of SNM). These drivers lead to the proposal of a compact, solid, highly-segmented,
cellular antineutrino detector using the ROL design developed for the LENS Collaboration’s solar neutrino experiment.

Earlier reactor antineutrino detectors  were deployed at distances over 10\,m from the core, with the most recent
experiments of Daya Bay, Double Chooz and RENO being $100+$ meters from the reactor cores (measuring
the $\theta_{13}$ antineutrino mixing angle at $1 - 1.5$\,km distance). All previous neutrino oscillation measurements were further from the
reactor cores than proposed for NuLat: $3 - 8$\,m (and $\sim$30\,m as explained later), and have been shielded with significant amounts of overburden, being underground at various depths.

A valid IBD produces a positron carrying most of the kinetic energy of the incoming neutrino, and a neutron receives most of the momentum.  Thus, measuring the kinetic energy of the positron yields a good measure of the neutrino energy.  The positron then rapidly annihilates with an electron at rest, forming two back-to-back 511\,keV gamma rays. In very large detectors, such as KamLAND or Borexino, the energy from the two 511\,keV gamma rays is integrated with the  positron energy and can be simply subtracted  from the  measured energy.  Unfortunately, in smaller detectors (on the ton scale), the gammas often deposit only a fraction of their energy before escaping the detector volume, thus adding large uncertainties to the deduced neutrino energy.  (NuLat segmentation avoids this problem by cleanly separating the positron and annihilation radiation.) The subsequent delayed-neutron capture then provides a powerful coincidence signal for background rejection. 

There are typically two approaches to addressing the background issues: shielding or discriminating event topology. Within mTC collaboration,  the latter
approach was tackled using first-light (defining  a so-called “Fermat surface”)~\cite{bib:mTC}  from interactions in a solid
scintillator, allowing millimeter reconstruction of events in a two-liter plastic scintillator detector. This is
orders of magnitude better than one would expect from a medium where the characteristic scintillation
decay time is many times the transit time of light across the detector. This was made possible by a
combination of very highly-segmented photon detectors (256 pixels per side), fast ($<$ 100\,ps) electronics, and because the time of the first-light ($t = 0$) is the most probable emission time from the exponential decay of scintillation light. Such event
topology could in principle allow location and timing of discrete energy deposits within the detector
resulting from a neutrino interaction (potentially including) the direction of the neutron recoil).
However, based on the 12.5\,cm cell results, without impractical numbers of pixels scaling up to a one-meter TimeCube it
was found to be impossible to reliably identify and reject the Compton scatters of the departing 511\,keV annihilation
gammas, and thus uncertainty on the positron energy and vertex location increased significantly.

The mTC collaboration considered a classic tubular array detector geometry, but concluded that the tubular
design also suffers the same problem as the open volume version: neutrino energy resolution suffering from
escaping annihilation radiation (and inability to practically locate the neutron capture). Moreover, in both of these schemes it was difficult to  contend with cosmic-ray
muons with little or no overburden (about 100/sec in a cubic meter detector), tails of extensive air
showers (EAS), energetic neutrons produced in the shielding and neighborhood which all contribute to
the trigger rate. A fast trigger, which recognizes a high total-energy deposition and/or high photodetector
multiplicity will eliminate a majority of these events, but may still allow an unacceptable rate leading to significant dead-time.

An example of a demonstrated tubular geometry detector is found in the 
Japanese PANDA Project~\cite{bib:PANDA}. They have operated a modest lattice (6x6) of tube 
style detectors (36\,m from the core) outside a \,3.4\,GW$_{th}$ power 
reactor in Japan, where they observed $21.8 \pm 11.4$ events /day in 
excess for reactor-on compared to $365 \pm 7$ events/day with reactor-off, 
for a S/N = 1/16.  Due to necessary cuts on the data prior to this, the 
efficiency however was a modest 3.15\%, making the net detected signal 
only 1/507 of neutrino-like triggers.  They predict a factor of 3 
improvement in efficiency with larger scale (10x10), however. \textit{This 
suffices to make one cautious about the background situation with such 
experiments.} This is of course directly relevant to other detectors 
operating closer to test reactors as in Table~\ref{table:exp_list}, where the 
1/distance$^2$ gain is offset by decreased power, and things are made 
worse by the direct reactor background, from which PANDA is shielded by 
distance and the reactor containment vessel.

High background trigger rates are generally addressed by detector segmentation: by making the region to be monitored for an event of interest as small as possible and utilizing the rest of the detector as veto volume.  In our case, the key is localizing the positron production and measuring the prompt signal energy for the prompt event. Using the nearby cells to register the delayed neutron capture, and constraining its association by time, space and energy separates an IBD event from the background. One can achieve this to a certain extent in the volumetric and tubular detector configurations, but the cellular design accomplishes the goal best, as discussed below.

0 developed by Raghavan's and Vogelaar's group at Virginia Tech for a solar neutrino experiment called the Low-Energy Neutrino Spectroscopy (LENS) experiment~\cite{bib:LENS}, which presented a much more challenging neutrino signal to background environment.

	In the course of studying the LENS design the VT team has built several versions of the lattice, and deployed a prototype (MicroLENS) underground in the Kimballton mine near Blacksburg, VA.  The LENS team at VT and collaborators at LSU and elsewhere, have made simulation programs and models, and studied the optics and overall detector response. The details of the ROL detector design as well as the details of previous constructions will be outlined below.

\subsection{Design Principles}
\label{sec:NuLatDesign}
\hskip 0.3in
		NuLat will employ a unique detector design, the ROL, as a means to separate the energy deposits of IBD positrons from the deposits of their annihilation gammas and to cleanly tag the delayed neutron to suppress random coincidences with the prompt signal. The ROL detector design uses complete 3-D optical segmentation, instead of typical 2-D segmentation and time-of-flight methods, for precision localization of events in a large volume detector. The segmentation of ROLs creates a detector composed of independent cells. Since the cells in a ROL detector are independent, short ranged events can be localized in a cell, thus the position resolution of a ROL detector is equal to the cell's size.  Furthermore, ROLs allow for the topology of multiple deposits to be analyzed. The optical 3-D segmentation of ROLs is accomplished using low-index of refraction barriers that allow for optical channeling of scintillation light down the primary coordinate axes. Since the lattice is constructed of a low-index of refraction material in a higher index medium, the channeling comes from total internal reflection (TIR) and Fresnel reflections. Figure \ref{fig:ROL} shows a simulation of this concept in 3-D.

		ROL detectors rely on the low-index barrier to channel light via TIR, but the choice of the barrier material depends on balancing several factors: generating a maximum of channeled light, generating a maximum of total light, and minimizing effects for varied propagation paths of the light. Depending on the specific indices of the bulk scintillator and the barrier, light will be: (1) channeled out directly along the primary Cartesian axes of the lattice into the PMTs that view the cell, (2) sprayed out into the three orthogonal planes containing the cell, (3) propagated out of the corners of the cell into other planes, or (4) trapped in the cell until absorbed by the scintillator. From these considerations ROLs are best designed with a minimum of trapped light  and a minimum of unchanneled light since increasing the trapped light reduces the total light collected and increasing the unchanneled light makes the analysis of the event topologies more complex. From a geometric analysis a $ \theta_{crit}  = 45^{\circ}$ provides 100\% light channeling with minimal light trapping, and a $ \theta_{crit} = 54.7^{\circ}$ provides 0\% light trapping with minimal unchanneled light. A second issue to consider is that as the percent difference of the scintillator index of refraction and barrier index of refraction increases, the Fresnel reflection at the scintillator-lattice interface increases. This reflection can become appreciable, on the order of a few percent at each lattice interface, and it grows proportionally to the number of cells in the lattice. These reflections affect the time structure of the pulses with a smaller critical angle giving more pulse time dispersion. 
		
\subsection{Previous Constructions}
\begin{figure}[!htbp] 
			\centering
		\includegraphics[scale=0.75, angle=0]{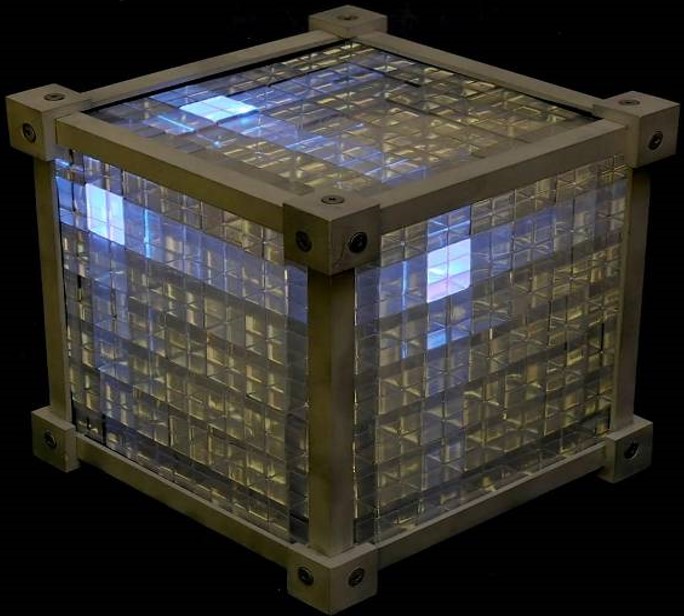} 
		\caption{Photo of a solid ROL demonstrator with light source position easily identified based on the light intensity radiating from the outside cubes.} 
		\label{fig:picoLENS}
	\end{figure}

	\hskip 0.3in
		Solid ROL detectors were conceived in the past, but never built past the demonstrator phase due to the higher cost of plastic scintillator for a large (100 ton) LENS scale detector. However, a solid ROL demonstrator was built previously by the VTech group and is shown in figure \ref{fig:picoLENS}.
		
\begin{figure}[!htbp] 
			\centering
		\includegraphics[scale=1.15, angle=0]{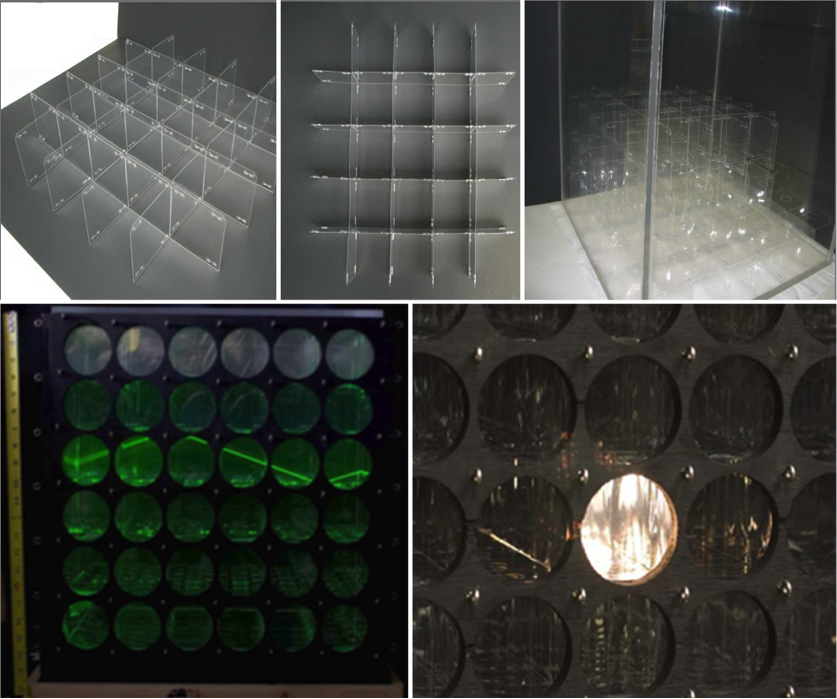} 
		\caption{Photo of a Teflon ROL demonstrator under construction (top), and filled with liquid (bottom) that illustrates the light channeling ability of an as-built ROL detector. The photo on the bottom left shows a side view of the lattice, illustrating the power of TIR using light from a laser beam. The photo on the right shows the end view of the ROL face with the bright light coming out of a single channel, transported via TIR from a source inside lattice, and almost no light is seen in other channels on the same lattice side.} 
		\label{fig:teflonLattice}
	\end{figure}

Development continued on designs segmenting liquid scintillator volumes with films or trapped gas  gaps. The most successful of these designs were ROLs made from Teflon FEP films (see figure \ref{fig:teflonLattice}). The microLENS detector and various ROL construction techniques for solid and liquid detection media are discussed in greater detail in the appendix.

\subsection{ROL Detector Construction for NuLat}
\hskip 0.3in		
	Due to the potential deployment of NuLat beside a naval reactor, safety concerns are an important consideration in the design. For this reason, a liquid scintillator based ROL detector would be problematic, but since the size of a reactor-based detector is much smaller, the higher cost of the plastic scintillator is not a large factor in the design. Therefore, a ROL made from a boron-loaded or (even better) a $^6$Li  loaded plastic scintillator will be used. This choice also simplifies the construction, since the plastic scintillator cells are easily machined and polished to be optically flat, and they can be stacked with small reflective spacers to ensure a low-index barrier is maintained without significantly affecting the light transport in the detector.
		
	The nominal design of NuLat is a lattice composed of 15$ \times $15$ \times $15 cubic array having 3375 cubes of EJ-254 boron-loaded (or as soon as available  Eljen $^6$Li-loaded \footnote{Eljen gives availability estimates as of early 2015.}) plastic scintillator 2.5" (6.35\,cm) on a side, and spaced 0.01" apart in . The base plastic for these is polyvinyl toluene (PVT), which has an index of refraction of $n_{scint}=1.58$ \cite{bib:EJ-254}. As described earlier, a gap index between 1.12 and 1.29 would be optimal for a PVT ROL, and NuLat has several options for the gap material that are nearly optimal. These include: air, perfluoroctane (PFO), and water, and their indices are roughly: 1.0, 1.255, and 1.33 respectively. For simplicity in construction the nominal design will use air gaps ($\sim$39$^{\circ}$ critical angle). Thus NuLat will have the vast majority of the light going to six PMTs that view the cell, with some light trapping and no unchanneled light.

\begin{figure}[!htbp]
\centering
\includegraphics[width=6.0in]{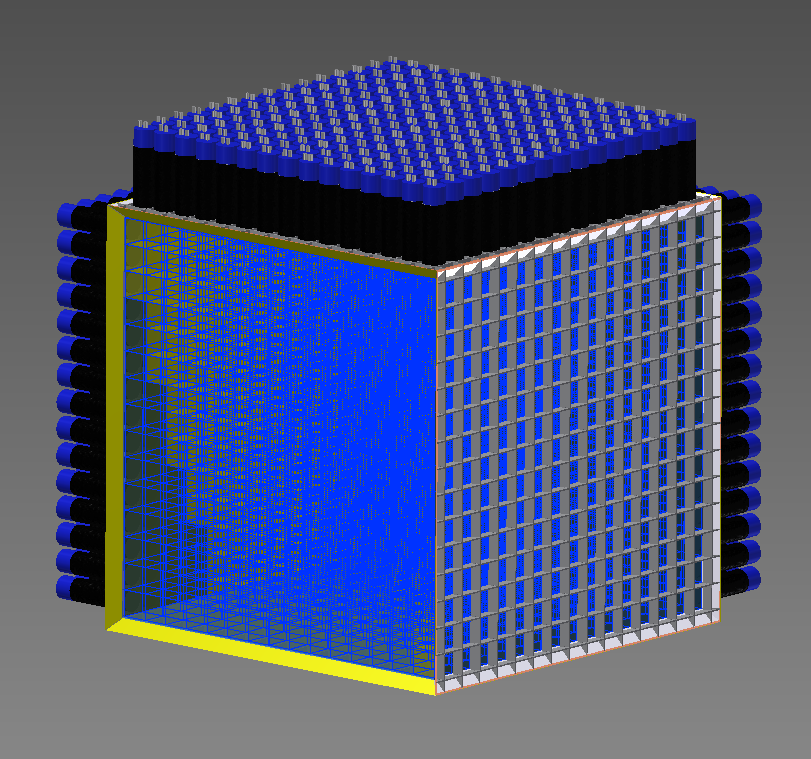}
\caption{CAD drawing of the NuLat 15$\times$15$\times$15 optical lattice with PMTs attached, allowing for 3D voxelization at the minimum resolution equal to the cell size. The PMTs have been removed from the three facing sides in the figure so that the  interior of the cube is partially exposed for better visualisation.}
\label{fig:NuLat_design}
\end{figure}

	NuLat will be instrumented with 2'' photomultiplier tubes (PMTs) connected to a short (1.5'' long) square channel to round photo-cathode light guide (LG) that will each view a single lattice channel. The LGs serve as an efficient means to couple the square cross section of the lattice channels to the circular photocathode. More detail on the use of LGs in ROLs can be found in the appendix. The full instrumentation will require a total of 1350 PMTs and LGs. Figure~\ref{fig:NuLat_design} shows a CAD drawing of the NuLat detector with PMTs attached (without LGs) and interior of the cube partially exposed for better visualization.

\section{NuLat Detector Simulation and Event Reconstruction}
\hskip 0.3in	
	To better understand the unique capabilities of the ROL detector and the NuLat experiment, the detector was simulated using the GEANT4 toolkit~\cite{bib:geant4}  and in-house light transport software. The result of these simulations and their impact on the reconstruction of the IBD positron’s energy are discussed below. 
	
\subsection{IBD Events in a NuLat Detector: Concepts}
	\hskip 0.3in
	For reactor sources of  IBD events, the positron will almost always deposit its energy in one cell, occasionally two cells, and very rarely in more than two cells. The annihilation gammas will typically scatter around the detector, and deposit their energy in a few cells around the vertex. This allows for the positron's kinetic energy to be reconstructed separately from the total energy of the prompt signal (i.e.\ the vertex cell contains mostly the energy from the positron with little contamination from the annihilation gammas). This is a unique feature of NuLat, and more specifics of positron events in NuLat will be detailed below. The neutron tag topology in the nominal NuLat detector is fairly simple since the IBD neutron will likely capture near the vertex cell, with a mean capture time of a few microseconds. For capture on $ ^{10} $B a small light pulse is seen in the capture cell from the 1.47\,MeV alpha and the 0.84\,MeV $ ^{7} $Li nucleus; 94\% of the time this is accompanied by a 477\,keV gamma  \cite{bib:scinNIST} from the de-excitation of the $ ^{7} $Li nucleus. The gamma will then deposit its energy in other cells as it scatters around the detector. On the other hand, neutron capture on $^{6}$Li produces a 2.05\,MeV alpha and a 2.73\,MeV triton which will produce signal in the capture cell of about 483\,keV$ _{\mathrm{ee}} $ making it well localized. Another aspect of the neutron tag is the average capture time between the prompt and delayed event. The capture time depends on the nuclei and level of loading in the scintillator, and this is especially important in the high background environment, where shorter capture times greatly improve background suppression of IBD. Outcome of the initial discussions with Eljen is that they can guarantee 1\% $^{10}$B loading with capture time of the order of 10 $\mu$s. 	
	
Finally, NuLat plastic test samples will be tested for the ability to discriminate neutron by pulse shape discrimination which has been demonstrated in $^6$Li doped plastic scintillator by Laurence Livermore National Laboratory (LLNL) group\cite{bib:Z}. Since this feature has not  yet been utilized  in plastics, NuLat baseline design does not rely on it. If proven successful, it would add additional neutron discriminating power to NuLat.  

	Other neutron capture nuclei such as Gd  have been proposed for very short baseline experiments, and its tag signature is noted here for completeness. Neutron capture on Gd produces about 8\,MeV of gammas with a multiplicity of $3 - 5$. While the total energy of the tag is well above possible backgrounds, the leakage of the gammas is potentially problematic in a small detector. This reduces the effectiveness of the tag. Regardless of the neutron capture nuclei, the ROL design provides a unique means to detect the neutron tag from IBD events because the capture is constrained to a very limited 5-dimensional phase space (in $x$, $y$, $z$, $t$, \& E).
	
\subsection{Simulation of Positron Events in NuLat: Concepts}
\hskip 0.3in
	The current simulations use the minimal NuLat design with three sides of the detector mirrored and the other three side instrumented. The mirroring reduces the number of PMTs, electronic channels, and the space required for the NuLat detector. For the optical properties of the scintillator, the nominal figures in the EJ-254 scintillator's data sheet \cite{bib:EJ-254} are adopted,  with an optical attenuation length of 3\,m. These numbers reflect our current best estimates for these parameters in NuLat. 
	
		 
	 
\subsection{Positron Event Reconstruction}
\hskip 0.3in
	This section will demonstrate the power of NuLat to measure the positron's energy with high resolution. The event reconstruction algorithm currently implemented uses only the PMT charges and the simulated detector non-uniformity to do a $ \chi^{2} $ minimization of the energy deposits in each cell of the NuLat detector. To illustrate the algorithm $ 10^{3} $ 2\,MeV positrons were simulated uniformly throughout the minimal NuLat discussed above.
	
	The event reconstruction algorithm starts with the PMT charges and determines, cell-by-cell, if there was possibly some energy in that cell. Requiring three orthogonal PMTs that view the cell to have light in them achieves this goal. Then using the simulated non-uniformity of the detector, the minimum energy deposit needed to produce the observed signal in the PMTs is logged in the cell. Once all the cells have been checked, the expected PMT charges from these deposits are determined using simulation data. A $ \chi^{2} $ is calculated from the sum of the squared residuals from the expected PMT charges and the observed charges. This is then minimized in a gradient search in energy over the cells that have energy deposits. 
	
\begin{figure}[!htbp] 
		\centering
		\includegraphics[scale=0.3, angle=0]{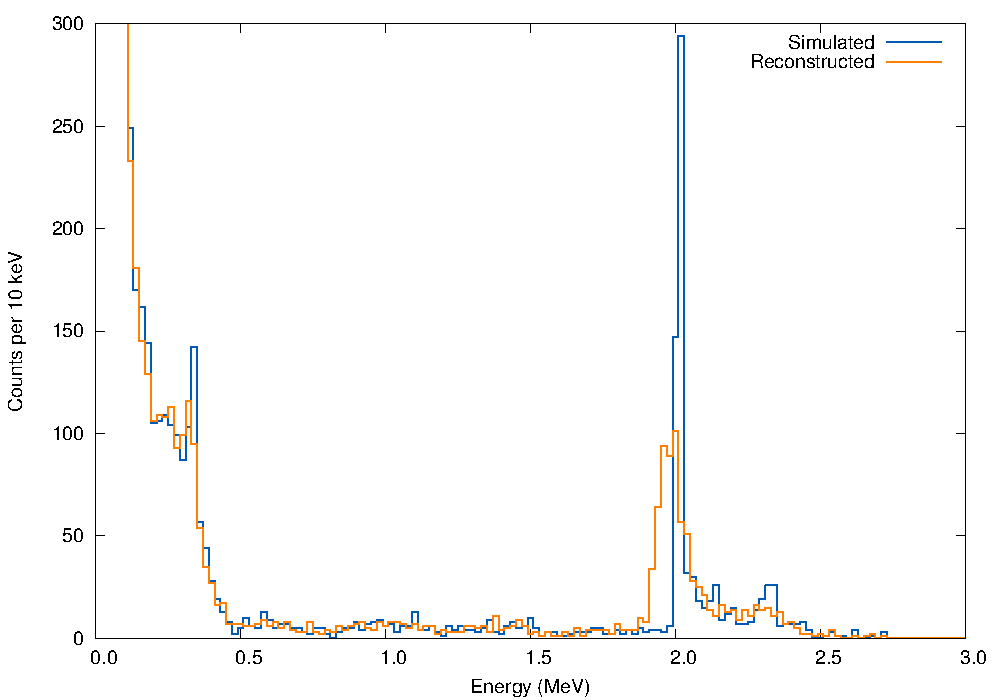} 
		\caption{The reconstruction of the response of individual cells within the detector due
to a 2\,MeV positron (and its subsequent annihilation) created randomly throughout the detector.} 
		\label{fig:dep}
	\end{figure}

\begin{figure}[!htbp] 
		\centering
		\includegraphics[scale=2.8, angle=0]{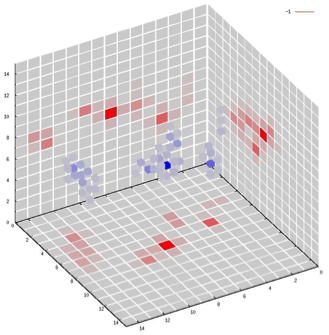} 
\caption{Reconstruction of a typical 2\,MeV
positron event. Red reflects PMT charge, and blue
reconstructed energy into cells. Note the large
energy deposit reconstructed into a single cell, and
the cloud of energy deposits into other cells due to
the two annihilation gammas.}
\label{fig:positron3D}
\end{figure}

	True energy deposits (blue) in the cells of the lattice are compared with the reconstructed cellular energy deposits (orange) in figure \ref{fig:dep}. Both histograms show the Compton edge for 511\,keV gammas at 340\,keV, the positron energy at 2\,MeV, and annihilation gamma contamination to the positron energy in the counts above 2\,MeV; the reconstruction gives a full width half maximum (FWHM) of $ \sim100 $\,keV at 2\,MeV. Figure~\ref{fig:positron3D} of a reconstructed 2\,MeV positron event. 
	
\begin{figure}[!htbp] 
		\centering
		\includegraphics[scale=0.4, angle=-90]{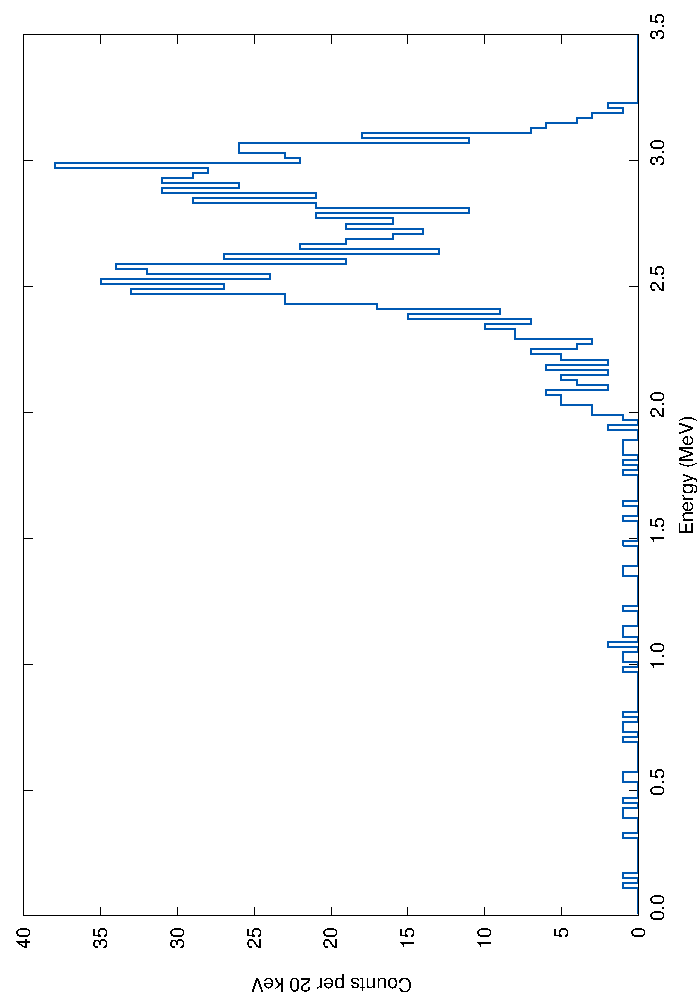} 
		\caption{The integral reconstructed energy of 2\,MeV positron distributed uniformly in a $ 15^{3} $ ROL detector, without using the cellular approach described in the text.} 
		\label{fig:totalReconEnergy}
	\end{figure}

	Next, the reconstruction of the total energy for 2\,MeV positrons stopping and annihilating  in the detector is shown in figure~\ref{fig:totalReconEnergy}. The total energy reconstructs in a range from about 1.9 to 3.1\,MeV. The majority of this broad width is due to leakage of the annihilation gamma's energy from the detector. However, this produces some interesting features in the spectrum. Peaked at about 3\,MeV is the full energy deposit for the positron and about 500 \,keV below this peak is another peak that corresponds to the escape of one of the 511\,keV annihilation gammas from the detector. There are not many events past the single escape peak because the back-to-back emission of the annihilation gammas makes the loss of both gammas unlikely; however, there is a small feature at 2\,MeV that corresponds to the double escape of the annihilation gammas. Past this, there are very few counts, which indicates how unlikely the positron is to escape from the detector.
	
\begin{figure}[!htbp] 
		\centering
		\includegraphics[scale=0.5]{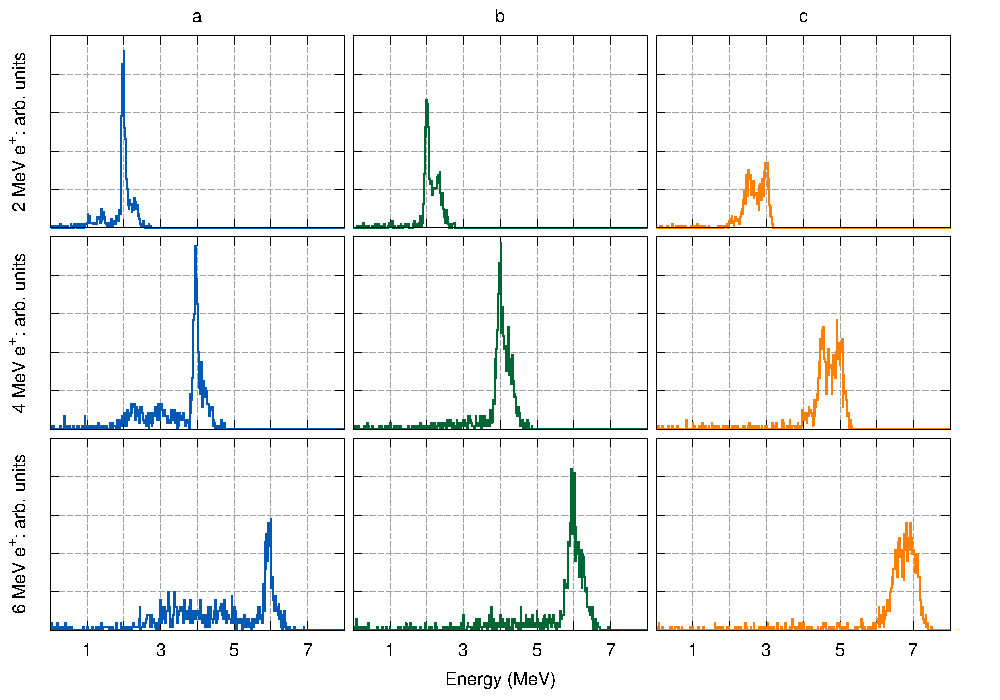} 
		\caption{The total reconstructed energy (c), the largest energy deposit (a), and the sum of the largest deposit and its largest neighbor deposit (b) for 2, 4, and 6\,MeV positrons distributed uniformly in a $ 15^{3} $ ROL detector.} 
		\label{fig:recon}
	\end{figure}

	The event reconstruction is not limited to the total energy deposit, because the spatial resolution afforded by the ROL detector design allows independent analysis of events on a cell-by-cell basis. Since the 2\,MeV positron will be contained in 1 to 2 cells and the annihilation gammas scatter and deposit their energy in several cells surrounding the vertex cell, the positron energy can be reconstructed more precisely by looking at the cell with the highest energy deposit and its largest deposit neighbor. These plots are shown in figure \ref{fig:recon}; column (a) (the blue histogram) is the energy of the cell with the largest reconstructed deposit; column (b) (the green histogram) is the combined energy of the cell with the largest energy deposit,  and its neighboring cell with the largest deposit; column (c) (the orange histogram) is the total reconstructed energy. In the plot, the 2\,MeV positron's reconstructed energy using only the largest deposits (blue) is slightly less than 2\,MeV, which is due to the leakage of the positron into an adjacent cell. Furthermore, there are more counts to the left of the peak relative to the plot in column b and less counts to the right. The counts to the right are events in which there is annihilation gamma contamination in the vertex cell. Therefore, inclusion of the largest neighbor cell to the vertex roughly doubles the annihilation gamma contamination.  It is clear by comparing the two histograms that the centroid of the reconstructed positron peak shifts toward higher energies, thus indicating that the positron's energy was fully contained in the two adjacent cells and the reconstruction gives a FWHM resolution of about 8\% at 2\,MeV. 
	
	Simulation of positrons at 4 and 6\,MeV further illustrates the need for the inclusion of the largest neighboring deposit to the vertex cell for the positron's energy determination. These plots are also shown in figure \ref{fig:recon} with the columns defined in the same way as above. As the positron energy increases it is more likely to deposit energy in a neighboring cell, and by comparing peaks one can see that the number of counts to the left of the peak in column (a), grows as the energy increases. These counts move to higher energies, as before, when the largest neighbor cell is included. 
	
\begin{figure}[!htbp] 
		\centering
		\includegraphics[scale=1.0]{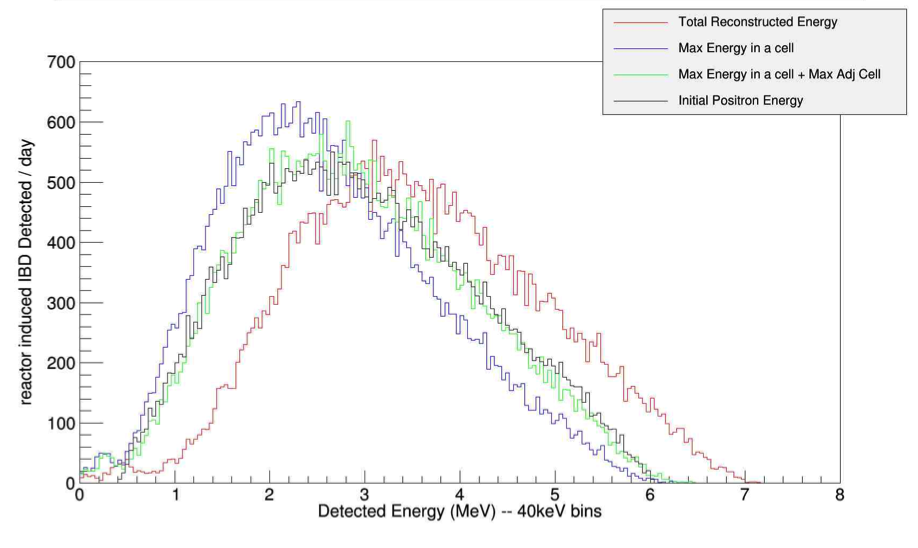} 
		\caption{The reconstructed positron energies in NuLat from simulated IBD positrons for one day at 3\,m from a 1.5\,GW$_{th}$ compact reactor compared to the initial positron energy.  The IBD spectrum is based on the G\"{o}sgen reactor experiments\cite{bib:gosgen}. Three cases are shown: 1) using the cell with the maximum energy deposition (blue), 2) using the cell with the maximum energy deposition plus the adjacent cell with the largest energy deposition (green), and 3) the total energy deposited in the detector, i.e. a traditional detector (red).  The initial positron energy is shown in black.  The detector response for case 2) tracks well the initial positron energy.} 
		\label{fig:pos_spectrum}
	\end{figure}

\begin{figure}[!htbp] 
		\centering
		\includegraphics[scale=1.0]{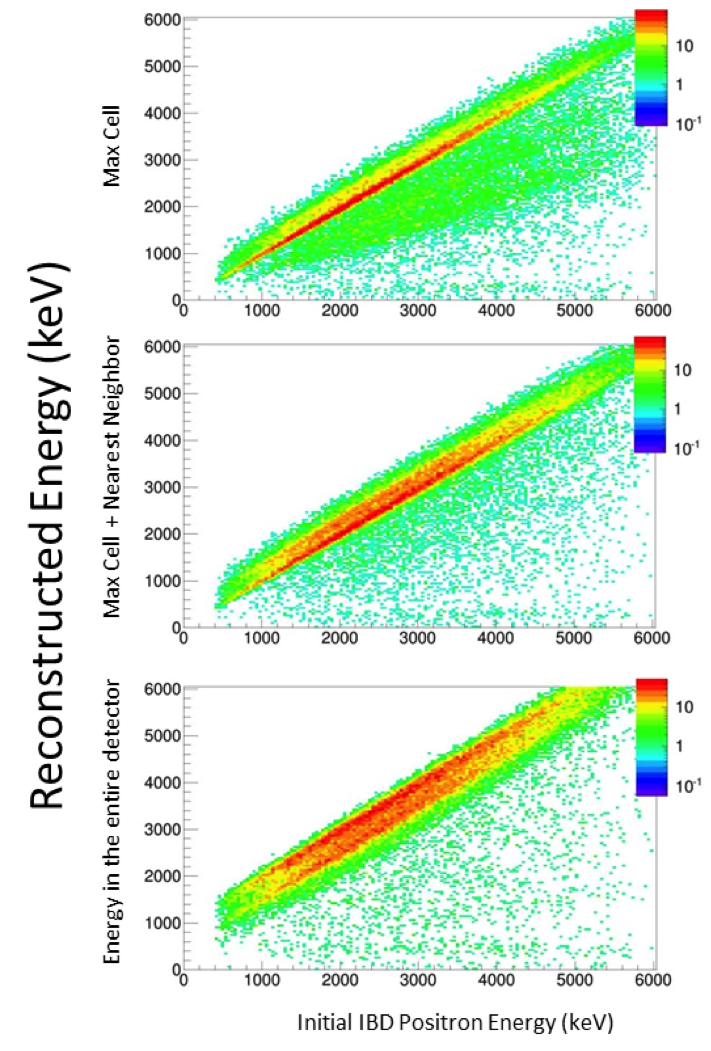} 
		\caption{Here are shown reconstructed positron energies in NuLat from simulated IBD positrons for one day at 3\,m from a 1.5\,GW$_{th}$ compact reactor compared to the initial positron energy.  The IBD spectrum is based on the G\"{o}sgen reactor experiments\cite{bib:gosgen}.  Three cases are shown: 1) using the cell with the maximum energy deposition (top), 2) using the cell with the maximum energy deposition plus the adjacent cell with the largest energy deposition (middle), and 3) the total energy deposited in the detector, i.e. a traditional detector (bottom).  Note the detector response for case 2) tracks the initial positron energy.} 
		\label{fig:ps2}
	\end{figure}

Lastly, the expected positron spectrum based on the G\"{o}sgen reactor experiments \cite{bib:gosgen}  is simulated with and without the best fit sterile oscillations in figure~\ref{fig:pos_spectrum} and figure~\ref{fig:ps2}. 
	
	NuLat is unique among the proposed very-short baseline experiments because of FWHM energy resolutions of $ \sim $8\% for positrons can be obtained. These tight energy resolutions are due to the ability of NuLat to separately reconstruct the positron's energy from the annihilation energy, which is afforded by the ROL segmentation.		
		
\section{Sensitivity to Sterile Neutrinos}
\begin{figure}[!htbp]
\centering
\begin{minipage}[h]{0.90\textwidth}
\includegraphics[scale=1.7]{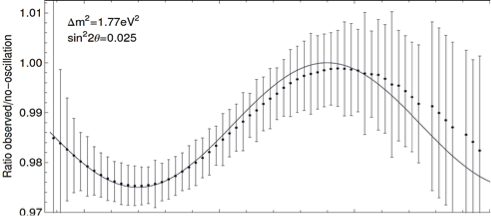}
\end{minipage}
\vskip 0.1in
\vfill
\begin{minipage}[h]{0.90\textwidth}
\includegraphics[scale = 1.7]{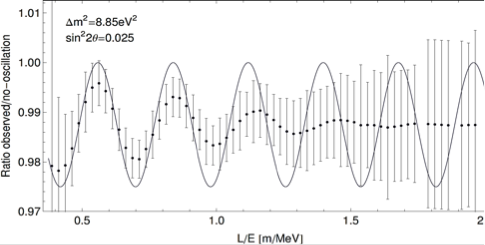} 
\end{minipage}
\caption{L/E oscillation pattern for two different choices of sterile neutrino mass difference assuming a point-like reactor neutrino source and include uncertainty in the spectral shape.}
\label{fig:LoverE}
\end{figure}

\begin{figure}[!htbp]
\centering
\includegraphics[width=5.0in]{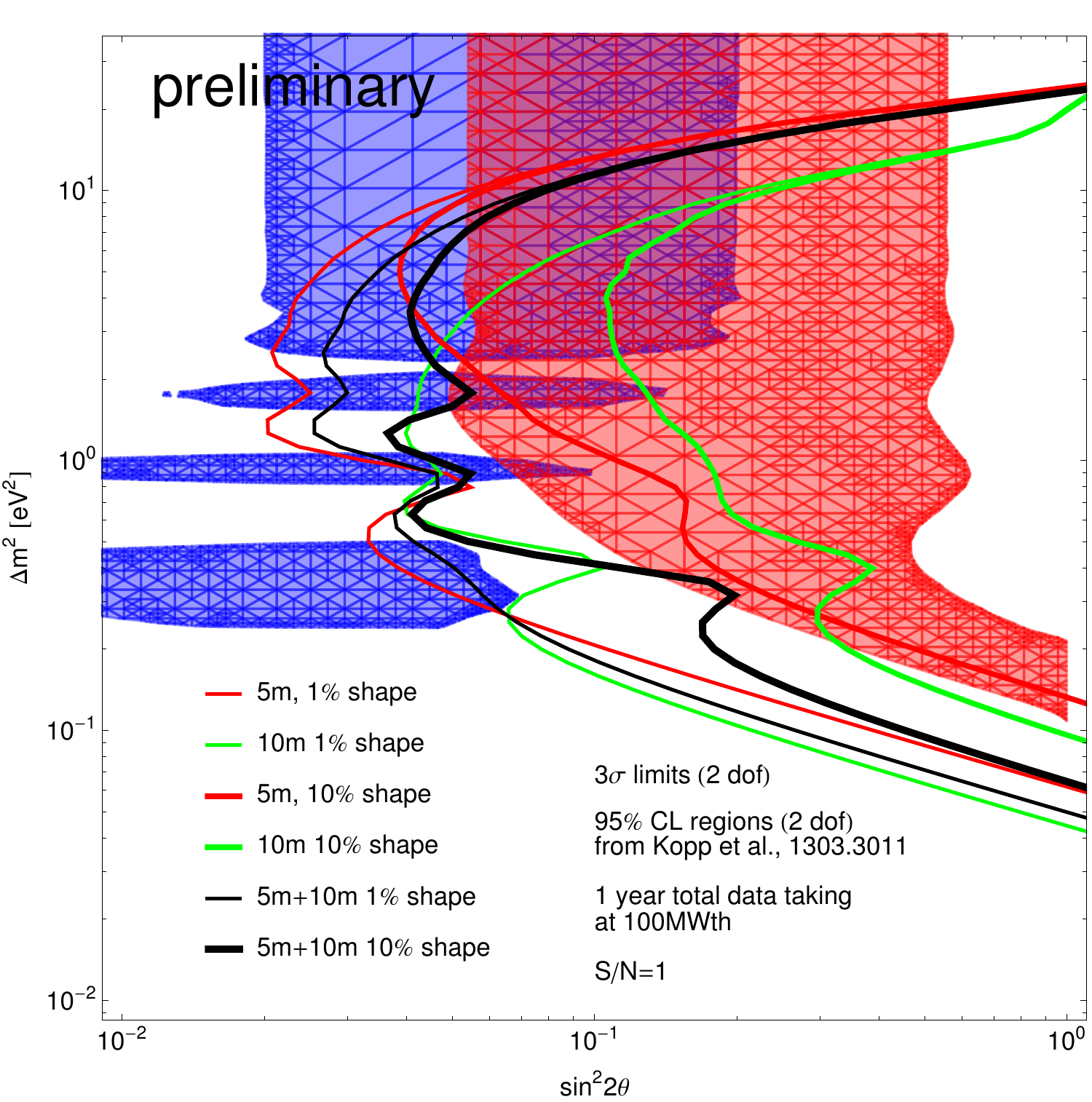}
\caption{NuLat sensitivity contours that correspond to 1 year exposure at 100\,MW$_{th}$ or 24 days/single station, or 12 days each/2 stations exposure at 1.5\,GW$_{th}$ US Navy reactor.}
\label{fig:sensitivity}
\end{figure}

\hskip 0.3in
Figure~\ref{fig:LoverE} shows oscillation pattern for different choices of sterile neutrino mass difference assuming point like reactor source. Excellent ability to observe oscillations even at close to 10\,eV$^2$ can be noted.  Figure~\ref{fig:sensitivity} shows the sensitivity of NuLat for testing RAA. Projected sensitivity is based on 1535 2.5'' cells and a 0.53\,m$^3$ cubic-reactor core with uniform
power density. Calculations were done 3.5 and/or 7.0\,m from the core, with 1\% or 10\% uncertainty
in the spectral shape. While the sensitivity is comparable to those of other short baseline reactor experiments (such as PROSPECT~\cite{bib:PROSPECT}), the clear advantage of NuLat originates from its unique lattice design that ensures a more powerful background rejection, vertex and energy resolution and a much shorter exposure time if operated at the US naval reactor of 1.5\,GW$_{th}$. While the sensitivity plot refers to a 1 year exposure  at 100\,MW$_{th}$ reactor, this translates to just 24 days and one station (or 12 days at each of two stations) in the vicinity of a 1500\,MW$_{th}$ US naval reactor. 

\subsection{Deployment Plan}
\hskip 0.3in
Two types of reactor sites have been identified for the NuLat experiment. The first type is a research reactor, in particular the NIST research reactor and the DOE Advanced Test Reactor (ATR) in Idaho. Their detailed features are described in~\cite{bib:Heeger}. While compact and fueled with $^{235}$U, they both have relatively low power, and being at the Earth's surface, very little shielding from reactor related backgrounds or cosmic rays.
The second type is a US naval reactor on the newest generation of US aircraft carriers that are very compact, well shielded and powerful: $\sim$1.5 GW$_{th}$. Because of the naval reactor compactness, NuLat can assess baselines between $3 - 6$\,m with a high neutrino flux.

The NuLat detector will be first deployed at NIST, ATR or possibly ORNL, for initial commissioning and data taking and then taken to a US Navy reactor for its final run. The final run is anticipated to take less than a year, allowing NuLat to collect a large data sample of IBD events, for a precise test of the RAA in a short time.

\section{Summary of NuLat Project}
\hskip 0.3in
The Neutrino Lattice design is unique among very short baseline reactor neutrino experiments in search of  evidence for sterile neutrinos, and will produce superior results in signal and removal of background due to the high level of 3D voxelization.  This detector has multiple other uses, including investigating the peculiar $5 - 6$\,MeV spectral bump seen in the RENO, Double Chooz and Daya Bay experiments, precise measurement of reactor neutrino flux, and being useful as a reactor monitor and as a neutron detector.

Most of the resources have been identified and are starting construction in early 2015, with the goal of testing at NIST in late 2015 and going to a USN reactor in 2016.

\section{Appendix A: ROL Design}

\subsection{MicroLENS}
\hskip 0.3in
	MicroLENS was a prototype ROL detector built by the LENS collaboration as a test bed for lattice construction, filling, and operating. The lattice for microLENS was made from 0.002" thick Teflon FEP strips and sheets. The strips were creased and then woven around 0.5 mm quartz rods. These strips defined the sides for a layer of cells in microLENS. The layers had a sheet of FEP placed above and below; these sheets defined the top and bottom of the lattice's cells. 
	
	Teflon FEP has an index of refraction of 1.34 compared to typical scintillator of $ \sim $1.5, thus giving a $ \theta_{crit}=63.3^{\circ} $. This means that the Teflon FEP lattice has significant spray light; however, it gives much more total light than a comparable air gap lattice. Given the greater light yield and relative ease of construction microLENS was built with an FEP lattice. Details available in~\cite{bib:BJ}\cite{bib:Derek}.
	
\subsection{Other ROL Construction Techniques}
\begin{figure}[!htbp]
\centering
\begin{minipage}[h]{0.90\textwidth}
\includegraphics[scale=1.0]{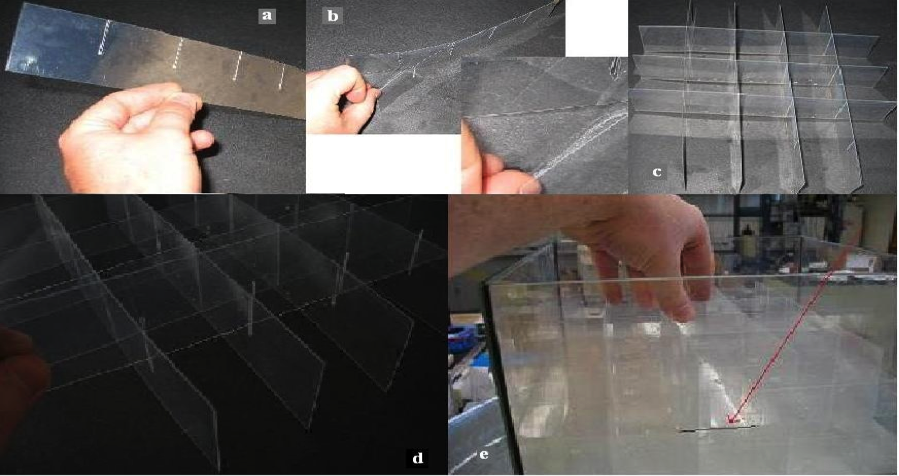}
\end{minipage}
\vskip 0.1in
\vfill
\begin{minipage}[h]{0.45\textwidth}
\includegraphics[scale = 1.0]{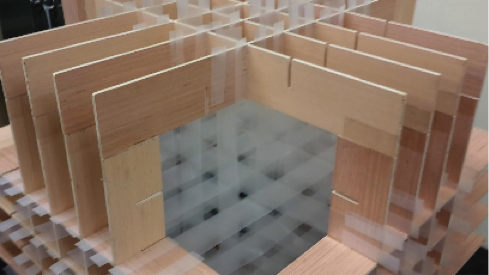} 
\end{minipage}
\hfill
\begin{minipage}[h]{0.45\textwidth}
\includegraphics[scale = 1.0]{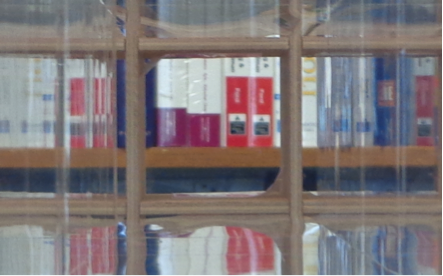} 
\end{minipage}
\caption{Techniques for building a large lattice detector with liquid scintillators.  (top a,b,c,d,and e) The primary building block of the “comb” construction techniques Teflon FEP on acrylic SL--  a) A strip of support structure material and Teflon FEP, with notches cut in it; b) The layering of the support structure material and the low index film;  c) Here the strips have been interlocked;  d) The last strip of one layer is being put in place;  e) The red arrow points to, the large sheet placed between the comb structures (the black line highlights a segment of this layer). Bottom left and right: Woven Teflon FEP strips construction techniques Teflon FEP SL. The lattice support framework (acrylic in practice) is pre-strung with translucent film strips (\textit{left}), used to pull through full-width 0.002'' thick transparent Teflon films, providing very tight tolerance cubes and excellent planarity (\textit{right}).}
\label{fig:microLENS}
\end{figure}

\hskip 0.3in
	The microLENS prototype was constructed using the strips of film, but this method is not suited to the construction of vapor gap ROLs. In addition to this, the microLENS construction method was time consuming due to the difficulty of working with such thin films and fragile quartz rods. Therefore, the LENS collaboration developed the slat method of construction. This method is essentially a 3-D weave of low index material that subdivides a volume into cells. The cells are defined by slats placed in the $ x $, $ y $, and $ z $ directions. The concept is illustrated in figure~\ref{fig:microLENS}.

	To use the slat construction method for a Teflon FEP ROL, a thin coating on a stiff acrylic plastic substrate was investigated, but the coating FEP onto acrylic is problematic due to the temperatures involved. An alternative to the coating is to couple a thin sheet of FEP to the acrylic backing; however, the resulting lattice strip's planarity was not sufficient and the concept was abandoned. The best method for producing a FEP lattice using the slat method is to simply use strips of Teflon FEP with tensioners to keep the films flat. 
	
	The slat construction method is also suited to the building of vapor gap ROLs. These ROLs can be made by taking two sheets of thin (0.015") acrylic and bonding with a frame of acrylic. This leaves a 0.015" vapor gap between the outer sheets of acrylic. The gap can be filled with the desired material\textemdash air or perfluorooctane (PFO)\textemdash for example. The drawback to this construction method is that since the slats are fairly thick they introduce significant lattice channel cross talk. This cross talk can be eliminated, but at the expense of the light yield. Moreover, for long-term deployment there is no guarantee that the integrity of the slats will be maintained. Details in ~\cite{bib:BJ}\cite{bib:Derek}.

\subsection{NuLat Light Guides (LGs)}
\label{sec:NuLatLightGuides}
\hskip 0.3in	
		LGs can have various roles in a ROL detector that depend on the detector's specific construction and deployment. However, for all ROL detectors LGs provide an efficient means of light collection from the lattice's square cross section to the round cross section of a PMT's photocathode. It should be noted that these LGs are especially efficient because the light from the lattice is primarily moving along the LG’s longitudinal axis; hence, their transport efficiency is not as affected by Liouville's theorem. 
		
		ROL detectors that employ Teflon FEP as the low index barrier in a liquid scintillator produce a sizeable fraction of light that is not channeled. In order to maintain the best position resolution possible, (at the expense of energy resolution), it is necessary to remove this unchanneled, or ``spray'' light. The simplest solution is to use an external light guide that has first a square cross section that matches the dimension of the lattice cells with an approximately 2 cell dimension length. The LG then transitions down to the round cross section of the PMT. The square section is optically coupled to Teflon, which is then optically coupled to a light absorbing material, so that the channeled light continues to be channeled while unchanneled light passes through the Teflon and is absorbed. The same effect can be produced, and space saved, if inserts of a similar construction to the square section of the LG are placed in the outer most layer of the lattice. However, to get the same reduction in spray light the insert must be sub-divided into smaller square cross sections. This is needed to ensure that any unchanneled light has at least one or two chances of being absorbed. The inserts are also problematic because they introduce extra light blocking and additional complications in event reconstruction. ROL detectors that use air for the low index barrier produce little spray light and thus do not require the specialized LGs just discussed. 
		
		LGs can be made of a variety of materials, and their construction can be optimized to the specific deployment of the detector. For example, if a shielding liquid, like water or mineral oil surrounds the detector, then LGs can be constructed from clear or reflective films that transport light via TIR or specular reflection. These LGs allow for a low mass, and hence higher radiopurity, solution for low background applications. In the case of a ROL detector deployed in air it is best to construct the LGs from solid acrylic bars. These LGs transport light via TIR and are easily machined to have a square-to-round transition. 
				
		VTech, over the past few years, has developed expertise in modeling, constructing, and testing LGs. A simulation package has been written for modeling LGs with various cross-sectional shapes, including the square-to-round transitions described above, and for various constructions made from solids or films. For testing LGs, VT has developed a well-characterized robotic light source that has a time-averaged behavior of an isotropic point source of light mimicking isotropic scintillation light, for realistic evaluation of the LG performance. 
		
\section {Appendix B: Detector Optical Performance}
\hskip 0.3in
There are several techniques with many possible materials to construct the NuLat optical lattice. Simulations of the light output and efficiencies of several of the material options for NuLat have been performed using the GEANT4 toolkit \cite{bib:geant4}. Some simulation results plus the pros and cons of representative designs are presented below. 

Before going into the simulation results, a preliminary discussion of total internal reflection and its application to the NuLat detector is warranted. There are several competing factors to detecting scintillation light in NuLat: generate a maximum of channeled light, generate a maximum of total light detected, and minimize effects for varied propagation paths of the scintillation light as it leaves the detector. Balancing these factors depends on what materials are used to construct NuLat.

The first issue considered is the channeling of light out of the detector. For scintillation light that is emitted isotropically from scintillator with index of refraction, $n_{scint}$, in a cubical lattice of index of refraction, $n_{lat}$, total internal reflection will occur if the angle from the normal of a plane, $\theta$, for each face of the cube is greater than $\theta_{c}= \sin^{-1}(n_{lat}/n_{scint})$. Assuming isotropically emitted scintillation light and total internal reflection for each face of a unit cell, the fraction of the scintillation light can be calculated: (1) channeled out directly along the axes of symmetry of the lattice into six PMTs that define the vertex cell, (2) channeled out in the three planes defined by a cell and not into the direct PMTs, (3) sprayed out of the corners of the cell into other planes, or (4) that which is trapped in a cell until reabsorbed by the scintillator itself. In order to get perfectly channeled light with no leakage into the plane or out of the corners of a cell the index of refraction ratio should be $(n_{lat}/n_{scint})=1/\sqrt{2}$. But this leads to $\sim 29\%$ of the light getting trapped in the detector. To have no trapped light requires an index of refraction ratio to be $(n_{lat}/n_{scint})=\sqrt{2/3}$. But some fraction of light is no longer channeled, but rather sprays out of the channel, making analysis of the event topology more difficult.

The second issue to consider is that as the difference between the scintillator and lattice index of refraction increases, the reflection at the scintillator-lattice interface, $R$, increases as $R = \left( \frac{n_{lat} - n_{scint}}{n_{lat}+n_{scint}} \right)^{2}$, for incident light perpendicular to the surface of the lattice. The effect becomes appreciable with increasing indices of refraction ratio - on the order of a few percent at each lattice interface. This effect also increases proportionally to the number of cells in the lattice. This light is not necessarily lost, but adds to delayed light.
 
Lastly, if some lattice material other than air is chosen and not all of the light is channeled, then the edge of NuLat behaves as a single cell of an optical lattice and this complicates the interpretation of the channeled scintillation light.

First, several of the possible NuLat design options were simulated a positron with an energy distribution from $0.1$\,keV to $7$\,MeV approximating the distribution found from the positron spectrum measured at G\"{o}sgen \cite{bib:gosgen} and generated in the center cell only. The positron proceeds to annihilate into two $511$\,keV $\gamma$-rays and this process is simulated in GEANT4. For the present discussion  the subsequent neutron capture was not simulated. The fact that events are generated in the center cell is not a limiting factor since it is easy to determine where the positron is created due to the light channeling of the NuLat detector. For all of the simulations in this section, scintillation absorption length of $2$ meters is assumed. This is shorter than for most of the actual materials, but it is instructive to compare the different materials of NuLat on a more equal footing in order to evaluate the lattice light propagation effectiveness. Increasing the absorption length increases the amount of light detected in every NuLat design. For the air gap and perfluorooctane (PFO) medium simulations, there is also a cut requiring no large energy detected in the outer layer of the NuLat detector, which increases the uniformity of light detected by the detector. This cut is not possible for the water filled lattice due to the spray of light out of the plane of detection. Lastly in these simulations, light guides have not been used. This reduces the channeled light collected by $\sim$40\%. More detail of the impact of the light guides is discussed in section \ref{sec:NuLatLightGuides}.

For comparison of all of the possible detector materials, two histograms are considered, each of which can be used to estimate the energy of the positron. First, large amounts of light emanating from a single cell were considered, more than can possibly be created by a single $\gamma$-ray energy deposition. This is referred to as a {\it large} amount of light in each of the histograms. The actual value of {\it large} depends upon each NuLat design and sets the minimum energy of a positron that can be detected in NuLat. The second histogram shows the total amount of light detected by the entire detector. These two plots form a basic evaluation of channeling and total light output of each particular NuLat setup. The large light histogram is a good way to directly estimate the positron kinetic energy, while the total light histogram is a good way to estimate the energy of the positron plus the two 511\,keV $\gamma$-rays, because the $\gamma$-ray energy is, likely to be detected in several cells instead of concentrated in a single cell. Subtracting this expected $1022$\,keV $\gamma$-ray energy from the total energy is another way to estimate the positron initial kinetic energy.

For all of the simulations a 15$\times$15$\times$15 lattice was chosen with (2.5'' or 6.35\,cm)$^{3}$ cells. The three sides of the detector were mirrored. The mirror reduces the number of PMT plus electronics and reduces the space required for the NuLat detector. The non-mirrored design can only exceed the performance of the mirror design, thus all the minimal requirements derived from the mirrored simulations will certainly hold for the non-mirrored case. And for all of the following simulations, the light guides were skipped and results are described in section \ref{sec:NuLatLightGuides}. This reduces the light detected by approximately $40\%$ and increases the resolution of the simulation results accordingly.

\subsection{Light Channeling in Solid Detection Medium}
\hskip 0.3in
The NuLat lattice made up of solid doped scintillator such as EJ-254 (which is the default design option) has been considered first. This is an off-the-shelf material with index of refraction of 1.58 and a density slightly over 1.02 g/cc depending on the exact boron concentration. Possible lattice optical barriers include air (n = 1.0), PFO (n = 1.255), and water (n = 1.33). For all of these solid NuLat designs a lattice gap of $0.01''$ is used.

Since the solid scintillator (EJ-254) and air gap optical barrier is the easiest to build and is the basis of the proposed experiment, thus a more detailed analysis than for the other design options has been considered. This includes more details of energy deposits per cell and a preliminary energy estimation technique of the positron initial energy.

\begin{enumerate}

\item {\bf Index of scintillator n$_{scint}$ = 1.58 (EJ-254) and index of optical barrier n$_{lat}$ = 1.0 (Air)}

The light channeling is the best for the air optical lattice, $\theta_{c}= \sin^{-1}(1.0/1.58)=0.69$ rad, which means all the light is channeled from the cell of energy deposit directly to the three PMTs that look at the cell. However, there is also a fair amount of trapped light. When compared to other design choices this setup has a large amount of back reflected light, with an $R \sim 5 \%$ at each cell lattice interface. The number of PEs detected in large PE PMT is shown in figure~\ref{fig:largePMTvsKE0_air}. The positron energy is easily determined down to approximately 600\,keV. In figure~\ref{fig:totalPEvsKE0_air}, even lower  values of positron energy can be determined, if the signal can be distinguished from the background. More detailed analysis of the signal is expected to result in better event reconstruction, but this simple event evaluation demonstrates the power of the segmentation of the NuLat detector.

\begin{figure}[!htbp]
\centering
\begin{minipage}[h]{0.45\textwidth}
\includegraphics[width=80mm]{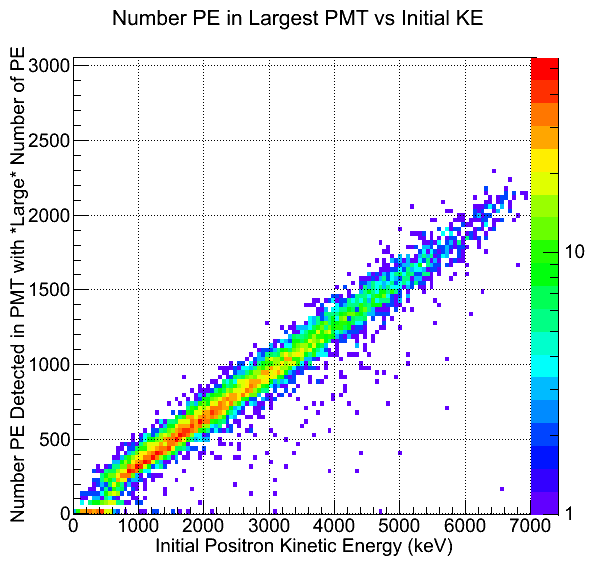}
\caption{Histogram of the light detected by all PMTs which detect more than 100 PEs per PMT vs. the positron initial kinetic energy. The histogram has been generated for EJ-254 solid with air filled gaps around each cell of the NuLat detector. Both figures have the same y-axis scale for easy comparison.} 
\label{fig:largePMTvsKE0_air}
\end{minipage}
\hfill
\begin{minipage}[h]{0.45\textwidth}
\includegraphics[width=80mm]{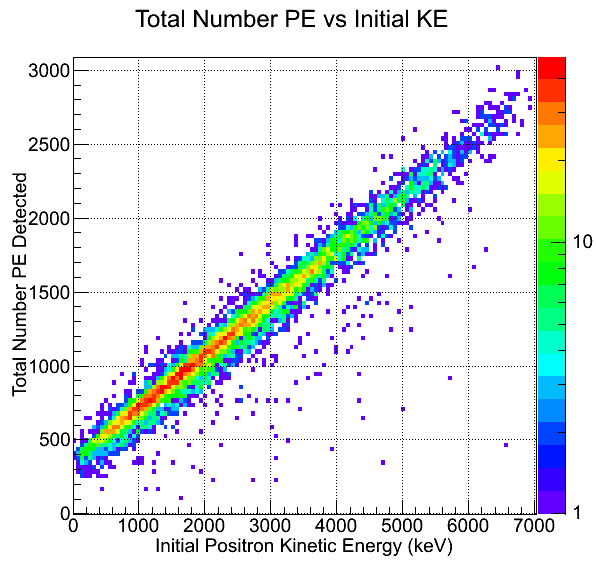}
\caption{Histogram of the total amount of light detected by all PMTs vs. the positron initial kinetic energy. The histogram has been generated for EJ-254 sold with air filled gaps around each cell of the NuLat detector. Both figures have the same y-axis scale for easy comparison.} 
\label{fig:totalPEvsKE0_air}
\end{minipage}
\end{figure}

A 2\,MeV positrons interaction with the NuLat detector was examined. First, the energy deposit per cell, both the recorded, as shown in figure~\ref{fig:recordededep}, and the detected optical energy (photo-electrons or PE) in a single PMT, as shown in figure~\ref{fig:pe} were analyzed. The particles that deposit energy are easily identified by the amount of energy deposited in individual NuLat cells. The low energy is energy deposited by the 511\,keV $\gamma$-rays in a single cell. The Compton edge at $\sim$340\,keV is clearly seen. All other energies are deposited either by the positron in one or more cells or by the positron and a $\gamma$-ray in a single cell. Selecting PMTs that see at least slightly more energy than the Compton edge for 511\,keV $\gamma$-rays produces a good measurement of the positron energy.  It can be seen from figure~\ref{fig:pe} that it corresponds to about 60 PEs for the solid scintillator cubes with air gap NuLat design option. 

\begin{figure}[!htbp]
\centering
\begin{minipage}[tl]{0.45\textwidth}
\includegraphics[width=80mm]{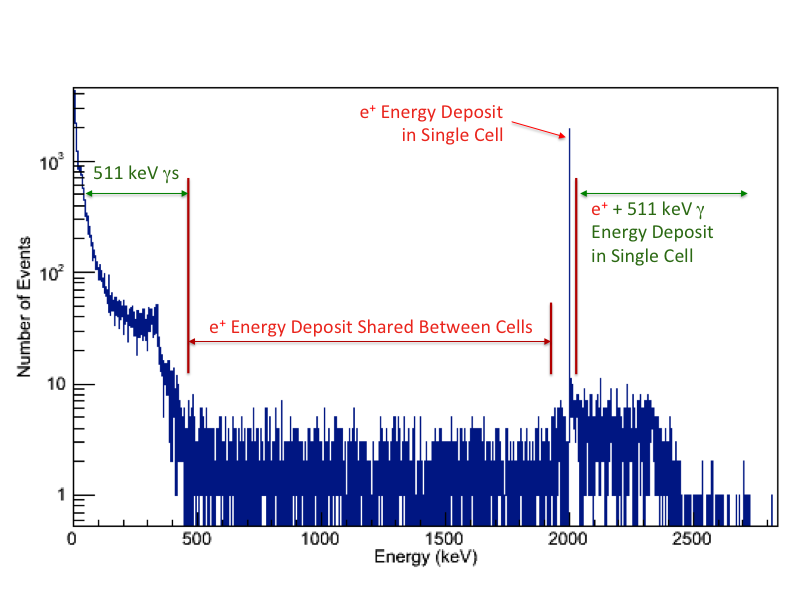}
\caption{Recorded energy per cell in the GEANT4 simulation. The 511\,keV $\gamma$ Compton edge is clearly visible at 340\,keV. The partial and full 2\,MeV positron energy deposits and the full positron energy deposit plus a 511\,keV $\gamma$ Compton edge are also clearly recognizable. The simulation has been performed for a single cell of EJ-254 solid scintillator with air filled gaps around each cell of the NuLat detector.} 
\label{fig:recordededep}
\end{minipage}
\hfill
\begin{minipage}[tr]{0.45\textwidth}
\includegraphics[width=80mm]{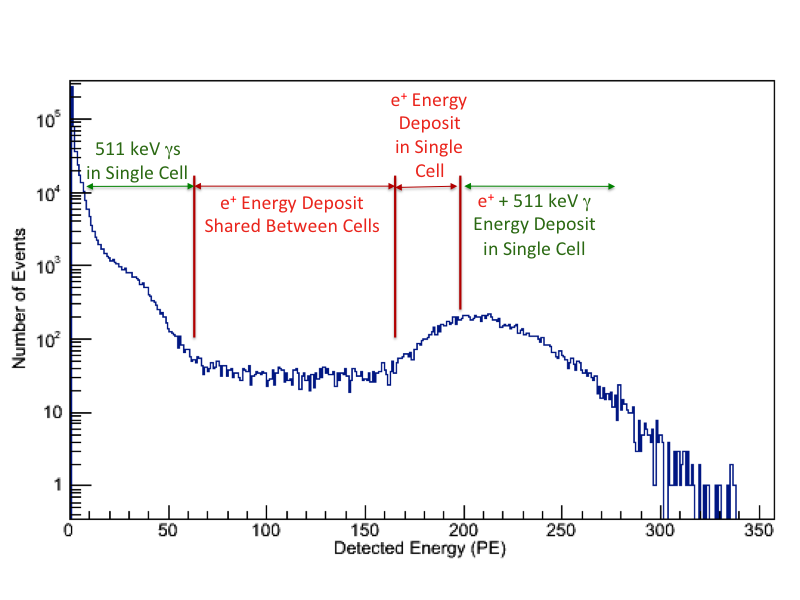}
\caption{The number of PEs detected in a PMT proportional to detected energy, for all individual PMTs. This graph is effectively the energy deposit shown in figure \ref{fig:recordededep} smeared by the single PMT energy resolution. The simulation has been performed for a single cell of EJ-254 solid scintillator with air filled gaps around each cell of the NuLat detector.} 
\label{fig:pe}
\end{minipage}
\end{figure}

A cut on the energy detected by a single PMT of greater than 60 PEs and sum all of these PMTs together for the 2\,MeV positron is shown in figure~\ref{fig:energyres2mevpos} in red. This positron energy estimate is referred as the ''large PE PMT energy estimate''. The full width half maximum (FWHM) of the reconstructed energy is about $17 \%$. When the light guides are modeled the energy resolution improves enough to separate out the Compton energy deposits in the same cell as the positron and hence improve the energy resolution and efficiency for the positron energy estimate. Another route to estimating the positron energy is to subtract the 1.022\,MeV $\gamma$-ray energy from the total energy deposit of an event. This plot for a 2\,MeV positron is also shown in figure~\ref{fig:energyres2mevpos} but in blue. This is referred to as  'total energy deposit' technique. It is interesting to note that the total energy deposit technique tends to slightly underestimate the positron energy (because of undetected $\gamma$-ray energy) while the large PE PMT method tends to slightly overestimate the positron energy (due to Compton scattering of $\gamma$-rays in the positron cell). These methods are complimentary to each other.

\begin{figure}[!htbp]
\centering
\includegraphics[width=120mm]{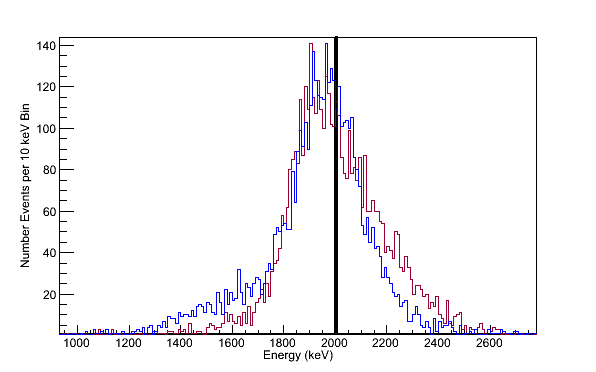}
\caption{Simulated energy deposition of a 2\,MeV positron in a single cell for the EJ-254 with air gaps around each cell of the NuLat detector. Recorded initial positron energy deposit (black), the large PE PMT reconstructed energy of the positron (red), and reconstructed energy of positron by subtracting out the 1.022\,MeV $\gamma$ energy from the total detected light (blue).} 
\label{fig:energyres2mevpos}
\end{figure} 

The last energy resolution item analyzed is the simulation of the full expected positron spectrum based on the G\"{o}sgen reactor experiments \cite{bib:gosgen} and the reconstruction of the positron energy based on the above basic algorithms. The results are shown in figure~\ref{fig:ibdspectrum}. Since this is a smooth spectrum and the positron energy resolution is at worst $15 - 20\%$\,FWHM, the evaluated spectrum looks promising. Using the large PE PMT method and ignoring background contamination, NuLat is capable of measuring positron energies down to $600$\,keV.

\begin{figure}[!htbp]
\centering
\includegraphics[width=120mm]{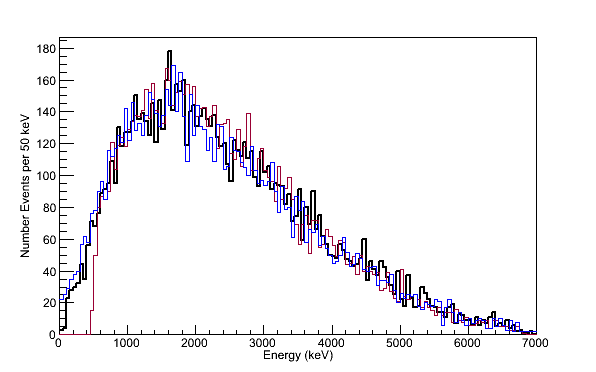}
\caption{Simulated energy deposits for inverse beta decay positrons in a single cell for the EJ-254 with air filled gaps around each cell of the NuLat detector. Recorded initial positron energy deposit (black), the large PE PMT reconstructed energy of the positron (red), and reconstructed energy of positron by subtracting out the 1.022\,MeV $\gamma$ energy from the total detected light (blue).} 
\label{fig:largePEPMT}
\end{figure}

One last interesting observation about the minimum number of PE detected is how many PMTs see that much energy per event. This is a direct measure of the positron's original location. If the light is completely channeled and the positron deposits energy in only one cell then only 3 PMT (6 in an non-mirrored detector design) should detect a large amount of energy. A complementary way to think about this is that each cell is effectively an independent detector that has 3 PMT watching it. But each PMT watches more than one cell so when two adjacent cells have large energy deposits only 5 PMT will see large amounts of light, not the naively expected $6$ PMT for the mirrored design. For a 2\,MeV positron the number of PMT that see more than 60 PE are shown in figure \ref{fig:largePEPMT} in red and the number of PMT that see any amount of light are shown in figure \ref{fig:largePEPMT} in blue. Notice it is most likely that either 3 or 5 PMT see large amounts of light. For higher energy positrons it is common to see 7 PMT that detect large amounts of light.

\begin{figure}[!htbp]
\centering
\includegraphics[width=120mm]{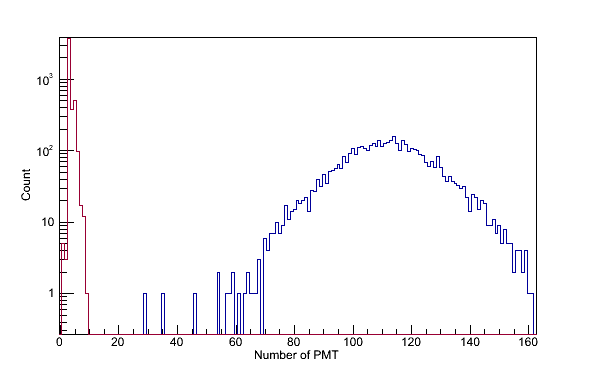}
\caption{The number of PMTs that detect light for 2\,MeV positrons for the EJ-254 with air filled gaps around each cell of the NuLat detector. The number of PMTs that detect a large amount of light (more than 60 PEs) are shown in red and the number of PMT that detect any amount of light are shown in blue. The large light PMT are related to the number of cells that the positron deposits energy, discussed in the text. } 
\label{fig:ibdspectrum}
\end{figure}


\item {\bf Index of scintillator n$_{scint}$ = 1.58 (EJ-254), index of optical barrier n$_{lat}$ 1.255 (PFO)}

For the PFO optical lattice, the best channeling is achieved, $\theta_{c}= \sin^{-1}(1.255/1.58)=0.92$ rad, but there is also light sprayed into the plane, however no light gets trapped. Comparing the total number of PE in both figure~\ref{fig:largePMTvsKE0_pfo} and figure~\ref{fig:totalPEvsKE0_pfo}, there is more light produced in the total PEs histogram per energy and in the large number of PEs PMT when compared to the air gap construction light output.

\begin{figure}[!htbp]
\centering
\begin{minipage}{0.45\textwidth}
\includegraphics[width=80mm]{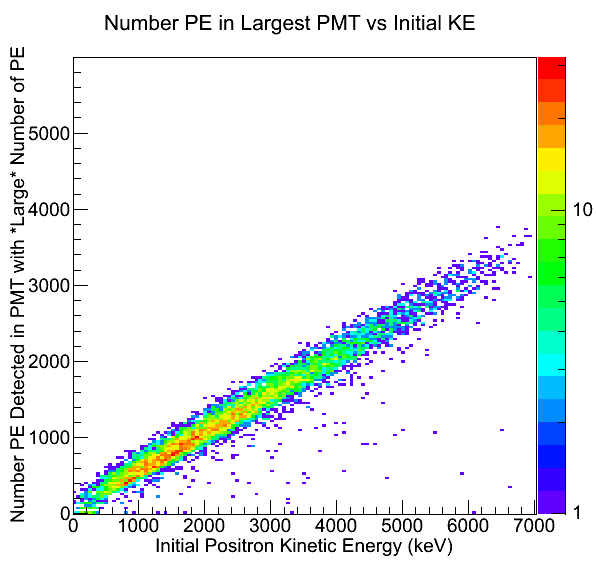}
\caption{Histogram of the light detected by all PMTs which detect more than 100 PEs per PMT vs. the positron initial kinetic energy. Histograms for EJ-254 solid with PFO filled gaps around each cell of the NuLat detector. Both figures have the same y-axis scale for easy comparison.} 
\label{fig:largePMTvsKE0_pfo}
\end{minipage}%
\hfill
\begin{minipage}{0.45\textwidth}
\includegraphics[width=80mm]{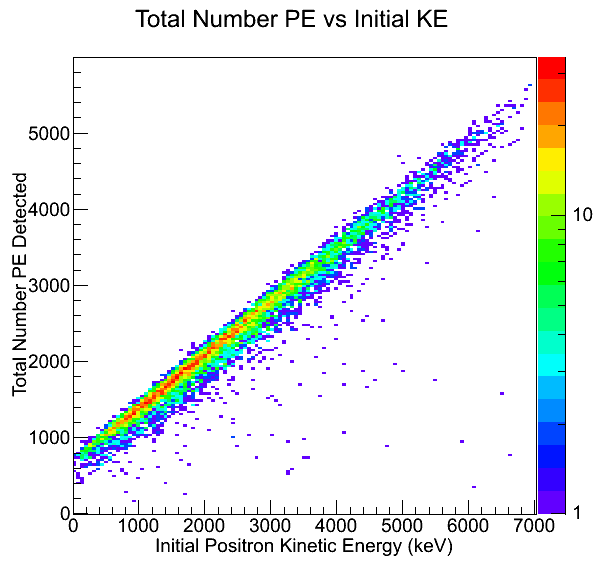}
\caption{Histogram of the total amount of light detected by all PMTs vs. the positron initial kinetic energy. Histogram generated for EJ-254 solid with PFO filled gaps around each cell of the NuLat detector. Both figures have the same y-axis scale for easy comparison.} 
\label{fig:totalPEvsKE0_pfo}
\end{minipage} 
\end{figure}

\item {\bf Index of scintillator n$_{scint}$ = 1.58 (EJ-254), index of optical barrier  n$_{lat}$ = 1.33 (Water)}

For the water optical lattice, the best channeling of all the solid scintillator options is achieved, $\theta_{c}= \sin^{-1}(1.33/1.58)=1.0$ rad,  but there is also light sprayed into the plane and into the corner PMTs of the detector. However no light gets trapped. Comparing the total number of PEs in both figure~\ref{fig:largePMTvsKE0_water} and figure~\ref{fig:totalPEvsKE0_water},  there is more light produced in the total PEs histogram per energy, but less in the large number of PEs PMT histogram when compared to the air gap construction light output.

\begin{figure}[!htbp]
\centering
\begin{minipage}[tl]{0.45\textwidth}
\includegraphics[width=80mm]{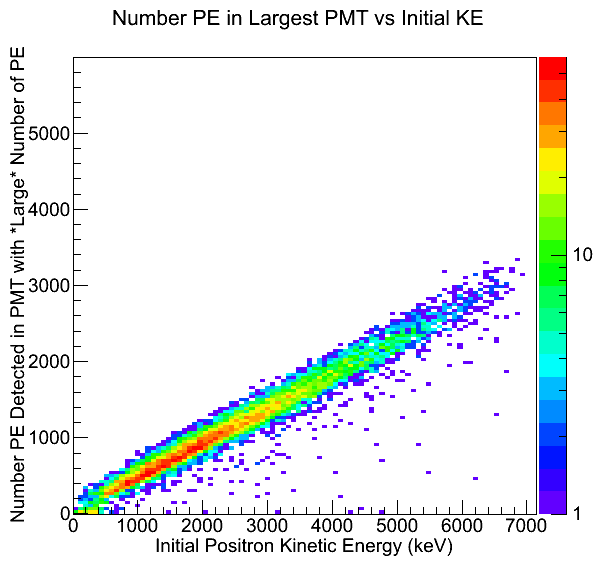}
\caption{Histogram of the light detected by all PMTs which detect more than 100 photo-electrons per PMT vs. the positron initial kinetic energy. Histograms for EJ-254 solid with water filled gaps around each cell of the NuLat detector. Both figures have the same y-axis scale for easy comparison.} 
\label{fig:largePMTvsKE0_water}
\end{minipage}%
\hfill
\begin{minipage}[tr]{0.45\textwidth}
\includegraphics[width=80mm]{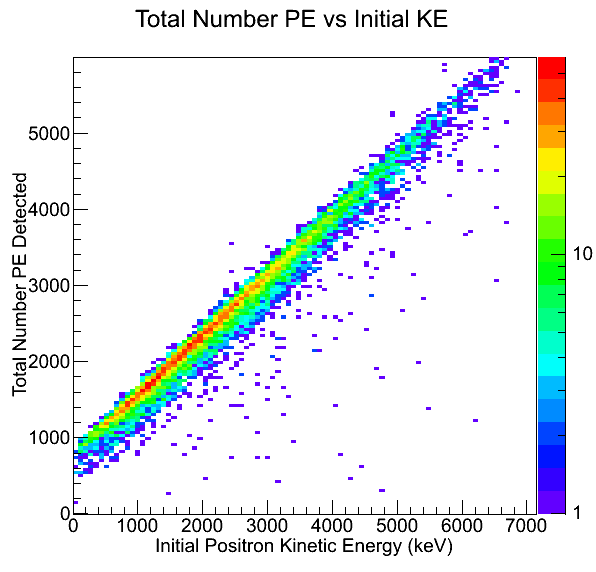}
\caption{Histogram of the total amount of light detected by all PMTs vs. the positron initial kinetic energy. Histograms for EJ-254 solid with water filled gaps around each cell of the NuLat detector. Both figures have the same y-axis scale for easy comparison.} 
\label{fig:totalPEvsKE0_water}
\end{minipage} 
\end{figure}

\begin{table}[!htbp]
\begin{center}

\begin{tabular}{ | c | c |}
\hline
Pros: & Ease of construction \\
\cline{2-2}
& Safety, no flammable liquids \\
\cline{2-2}
 & Complete channeling of optical light (Air) \\
\cline{2-2}
 & Up to $\sim10\%$ more dense than liquid scintillator \\
\hline
Cons: & Large back reflection at cell interfaces (Air)\\
\cline{2-2}
& Cost \\
\cline{2-2}
& Reduced total light collected (Air) \\
\cline{2-2}
& Reduced channeling of optical light (Water, PFO) \\
\cline{2-2}
& Edge effects (Water, PFO) \\
\hline
\end{tabular}
\label{table:solid_procon}

\caption{Pros and cons of the solid scintillator NuLat detector design.}

\end{center}
\end{table}

\end{enumerate}

\subsection {Light Channeling in Liquid Detection Medium with Single and/or Double Layer Barrier NuLat}
\hskip 0.3in
The basic doped liquid scintillator such as EJ-339 is considered next. This off the shelf material has an index of refraction of 1.415 and a density slightly over 0.92 g/cc depending on the doping concentration. Possible lattice optical barriers include air (n = 1.0), PFO (n = 1.255) and Teflon (n = 1.34). The air gap is considered only, as the others have much reduced channeling of light because of the lower index of refraction of the scintillator. Construction of the lattice for liquid scintillators is challenging and even the best construction techniques have measurable light leakage into the adjacent channels. For all of the liquid setups a lattice gap of $2 \times 0.381''$ acrylic and $0.0127''$ air gap is used.

\begin{enumerate}
\item I{\bf ndex of detection medium n$_{scint}$ = 1.415 (EJ-339), index of optical barrier  n$_{lat}$ = 1.0 (Air)}

For the air optical lattice, the best channeling is obtained, $\theta_{c}= \sin^{-1}(1.0/1.415)=0.78$ rad, which means light is channeled down the lattice and very little into the plane. Comparing the total number of PEs in both figure~\ref{fig:largePMTvsKE0_liqscint_air} and figure~\ref{fig:totalPEvsKE0_liqscint_air}, there is more light produced in the total PEs histogram per energy but less in the large number of PEs PMT histogram when compared to the air gap construction light output.

\begin{figure}[!htbp]
\centering
\begin{minipage}[tl]{0.45\textwidth}
\includegraphics[width=80mm]{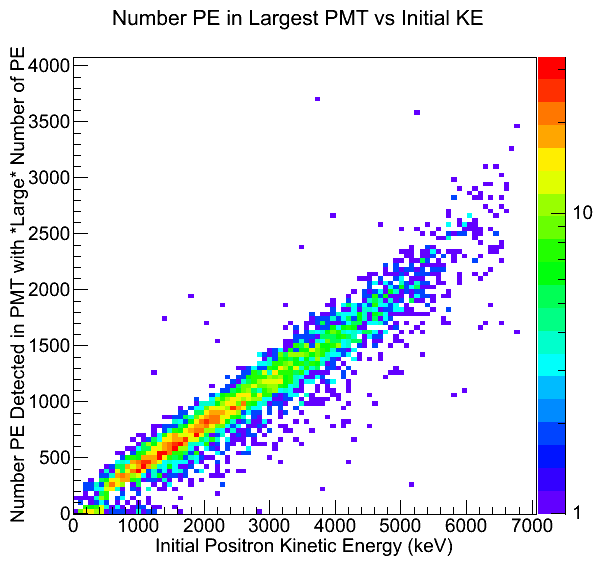}
\caption{Histogram of the light detected by all PMTs which detect more than 100 PEs per PMT vs. the positron initial kinetic energy. Histogram for EJ-339 liquid with air filled gaps between acrylic layers around each cell of the NuLat detector. Both figures have the same y-axis scale for easy comparison.} 
\label{fig:largePMTvsKE0_liqscint_air}
\end{minipage}%
\hfill
\begin{minipage}[tr]{0.45\textwidth}
\includegraphics[width=80mm]{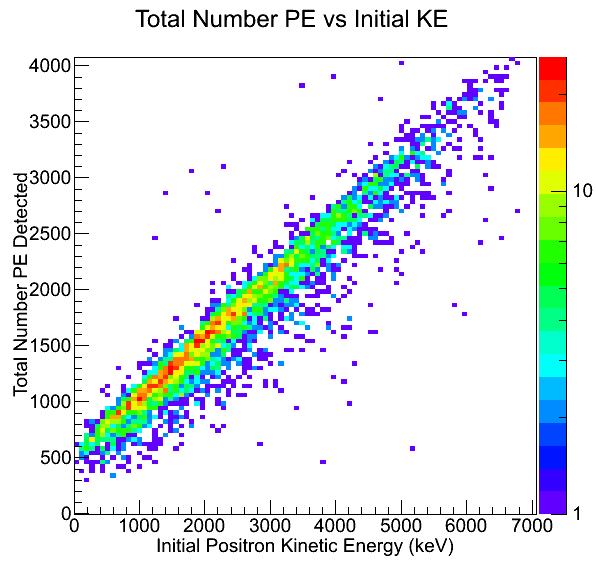}
\caption{Histogram of the total amount of light detected by all PMTs vs. the positron initial kinetic energy. Histograms for EJ-339 liquid with air filled gaps between acrylic layers around each cell of the NuLat detector. Both figures have the same y-axis scale for easy comparison.} 
\label{fig:totalPEvsKE0_liqscint_air}
\end{minipage} 
\end{figure}

\begin{table}[htp]
\caption{Pros and cons of the liquid scintillator NuLat detector design.}
\begin{center}
\begin{tabular}{ | c | c |}
\hline
Pros: & Cost of liquid scintillator \\
\cline{2-2}
 & Almost optimal channeling of optical light (Air) \\
\cline{2-2}
 & Construction already demonstrated with Teflon \\
 & and LAB scintillator ($n=1.54$) \cite{bib:LENSteflonplanarref} \\
\cline{2-2}
 & Increased total light collected compared to Solid+Air NuLat\\
\hline
Cons: & Liquids more difficult and more costly to handle, flammable \\
\cline{2-2}
& Construction of lattice with gas interface challenging to build \\
\cline{2-2}
& Reduced channeling of light (PFO, Teflon) \\
\cline{2-2}
& Short Absorption Length (Teflon) \\
\cline{2-2}
& Edge effects (Water, PFO, Teflon) \\
\hline
\end{tabular}
\label{table:liquid_procon}
\end{center}
\end{table}

\end{enumerate}

\subsection{Simulation outcomes}
\hskip 0.3in
Initial simulations of 2\,MeV positrons show that both solid and liquid scintillators and different barrier choices have their advantages and disadvantages. Solid detectors do not suffer from sprayed light and therefore have the most digital like signature. And while the overall amount of light reaching PMTs is less, there is still enough light to detect 600\,keV positrons, below the IBD threshold. Even with a very simple energy reconstruction, NuLat can easily achieve 15\% to 20\% energy resolution at FWHM (

\end{document}